\documentclass[%
preprint,
 amsmath,amssymb,
]{revtex4-2}

\usepackage{graphicx}% Include figure files
\usepackage{dcolumn}% Align table columns on decimal point
\usepackage{bm}% bold math
\usepackage{hyperref}% add hypertext capabilities
\begin{document}

%\preprint{APS/123-QED}

\title{Perfect optical spin-filtering in antiferromagnetic stanene nanoribbons induced by band bending and uniaxial strain  }% Force line breaks with \\

\author{F. Rahimi}
\email[Corresponding author's Email: ]{fatemeh.rahimi@azaruniv.ac.ir}

\author{A. Phirouznia}%

\affiliation{%
 Department of Physics, Azarbaijan Shahid Madani University, 53714-161, Tabriz, Iran\\Condensed Matter Computational Research Lab, Azarbaijan Shahid Madani University, 53714-161, Tabriz, Iran
}%

\date{\today}% It is always \today, today,

\begin{abstract}
	Non-equilibrium spin-polarized transport properties of antiferromagnetic stanene  nanoribbons are theoretically  studied under the combining effect of a normal electric field and linearly polarized irradiation  based on the tight-binding model at room temperature. Due to the existence of spin-orbit coupling in stanene lattice, applying normal electric field  leads to splitting of   band  degeneracy of spin-resolved energy levels in conduction and valence bands. Furthermore, unequivalent absorption of the polarized photons at two valleys which is attributed to an antiferromagnetic exchange field results in unequal spin-polarized photocurrent for spin-up and spin-down components. Interestingly, in the presence of band bending which has been induced by edge potentials,  an  allowable quantum efficiency  occurs over a wider wavelength region of the incident light. It is especially important that the variation of an exchange magnetic field generates spin semi-conducting behavior in the bended band structure. Moreover, it is shown that optical spin-filtering effect is obtained under the simultaneous effect of uniaxial strain and narrow edge potential.
	
\end{abstract}

\maketitle

%\tableofcontents

\section*{Introduction}

Recently, spin-functionalized optoelectronics properties of nanostructures have attracted significant interest in the scientific community~\cite{koppens2014photodetectors, ganguly2017magnetic, xiao2021spin}. In this field, interaction among the light and electrons by considering their spin degree of freedom in the absence of an external bias is studied. In modern optoelectronics, a large number of theoretical and experimental  works have been performed by considering optoelectronic and spintronic properties, simultaneously~\cite{endres2013demonstration, vzutic2002spin, ostovari2014dual}. Several approaches have been proposed to produce spin current using optical absorption~\cite{bottegoni2014spin, ellsworth2016photo}.

Successful synthesis and peculiar optical, electrical and spin transport features of graphene paved the way for utilizing of other two dimensional $ (2D) $ graphene-like materials in  optoelectronics applications. General name for $ 2D $ buckled monolayer structures which are formed by other group-IV elements is $ X $enes (silicene (Si-based), germanene (Ge-based) and stanene (Sn-based)). The stable structure of the buckled $ X $enes in comparison with graphene is a consequence of a $ sp^{2}-sp^{3} $ mixed  hybridization of Si, Ge, or Sn atoms in the forming process of these materials~\cite{takeda1994theoretical, cahangirov2009two}. The  layer separation between two triangular sub-lattices in this group of $ 2D $ materials is called buckling height~\cite{molle2017buckled}. This buckled structure allows us to tune  the band gap size by applying a vertical electric field~\cite{qi2018electric, ni2012tunable, drummond2012electrically}. One of the outstanding materials in this $ 2D $ material family is stanene. The existence of  strong spin -orbit coupling (arising from heavy Sn atoms) introduces stanene as an ideal $ 2D $ topological insulator.  Topological insulators are featured with an insulating band gap in the bulk and conducting state at the edges.  These phases are proposed as preeminent candidates for spintronic applications~\cite{pesin2012spintronics}. Recently, ultrathin Sn films with $ 2D $ stanene structure have been observed on the substrate of $  Bi_{2}Te_{3} $ by molecular beam epitaxy experiment~\cite{zhu2015epitaxial}. Also, other dominant physics features such as large-gap $ 2D $ quantum spin Hall  states~\cite{xu2013large}, giant magnetoresistance, perfect spin filter~\cite{rachel2014giant} and near-room-temperature quantum anomalous Hall effect~\cite{wu2014prediction} have been reported for stanene. 

Besides spintronic applications, stanene nanostructures are shown to be promising candidates for optoelectronic devices, and there have been important advances in this field during the last decade. Recently, electromagnetic response of staggered $ 2D $ lattices is studied via an external electrostatic field and circularly polarized laser light. It has been found that different topological phases in these lattices exert influence on the resonant behavior of  nonlocal Hall conductivity, considerably~\cite{rodriguez2018nonlocal}. Using density functional calculations,  electronic and optical properties of the graphene/stanene heterobilayers are investigated.  Combination of  the high carrier mobility of graphene and  excellent spin Hall effect of stanene in these heterobilayers  facilitate the performance of stanene related spintronic devices and suggest this material as a suitable candidate for photoelectronic devices~\cite{chen2016electronic}. In a previous study using  work function computations, it is revealed  that stanene and doped stanene have lower
work function in comparison with graphene and  accordingly stanene is a good candidate for photocatalysis devices \cite{garg2017band}.

Mechanical control  or so-called strain engineering  is reported as a suitable procedure for tuning  optical and electronic  properties in $ 2D $ allotrope of group-IV elements~\cite{levy2010strain, pereira2009strain,  roldan2015strain, guinea2010energy}. Due to the special structure of stanene, it  seems  possible that  one can tune stanene’s electronic properties by applying external strains. In this regard, some studies have been performed. By taking into account many-body effects, optical properties of stanene and stanane (fully hydrogenated stanene) have been investigated by applying strain.  Strain induced  optical gap of stanane provides valuable information on the potential application of stanene optoelectronic devices such as solar cells~\cite{lu2017quasiparticle}. Furthermore, continuous evolution of the electronic bands of stanene  across  nanoribbon is reported under the effect of  strain field gradient~\cite{lu2017quasiparticle}.

Besides the strain, there have been performed numerous studies on the band structures engineering and conductance properties of $ 2D $ materials. The edge potential is regarded as an efficient approach for engineering of electronic structure. Specially, Rachel and Ezawa observed quantum spin Hall effect without edge states only by employing different perturbations at the edges of silicene nanoribbons~\cite{rachel2014giant}.

Inspired by the increasing attention paid to spintronic and optoelectronic, in this study spin-dependent optoelectronic properties of antiferromagnetic zigzag stanene nanoribbons $ (ZSNRs) $ are investigated theoretically at room temperature.  Electric field spin-polarized photocurrent is computed by means of the self-consistent nonequilibrium Green’s function $ (NEGF) $ formalism and the tight-binding Hamiltonian under the effect of a linearly polarized illumination. In order to improve the performance of the system, effects of some external fields, such as edge potential and uniaxial strain are studied. The results show that applying a vertical electric field at two edges of the antiferromagnetic device results in band bending effect in the corresponding spin-resolved band structure of $ ZSNR $. It was shown that with the narrowing of edge potentials, one could broaden the acceptable photoresponsivity in a broad range of incident photon energies as well as increase  optical spin polarization percent. Finally, the effect of uniaxial strain on the optical spin transport properties of the antiferromagnetic nanodevice in the presence of edge potential is studied. The numerical results reveal nearly full optical spin-filtering properties under the effect of different strains.

 \section*{Results and Discussion} 
  {\label{sec:level2}Spin-polarized photocurrent}\\
Spin-photocurrent across the antiferromagnetic single layer zigzag stanene  nanoribbons was simulated through the tight-binding  approximation and the $ NEGF $ formalism by considering the combining effect of the normal electric field and the linearly polarized light. The incident light is monochromatic with constant intensity of $ I_{w}=100\,\frac{kW}{cm^{2}} $, which is radiated normally on the top of  central channel. In this study,  the scattering region has constant length which consisting of $ 120 $ unit cells which is sandwiched between two semi-infinite  left and right leads. In addition, the scattering region, the left and  right leads have  same structure.

\begin{figure*}
	\begin{center}
		\includegraphics[height=7cm]{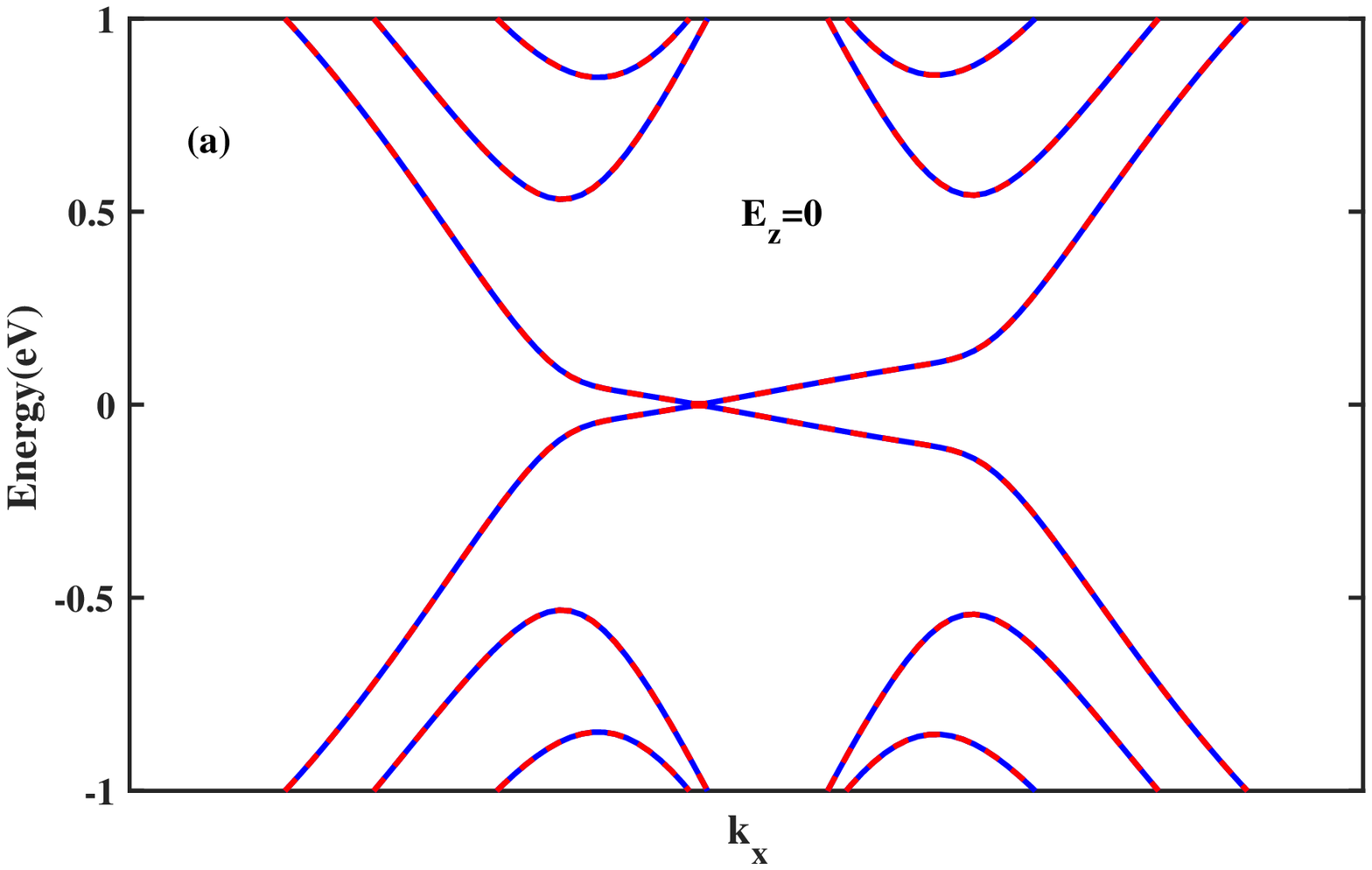}
		\includegraphics[height=7cm]{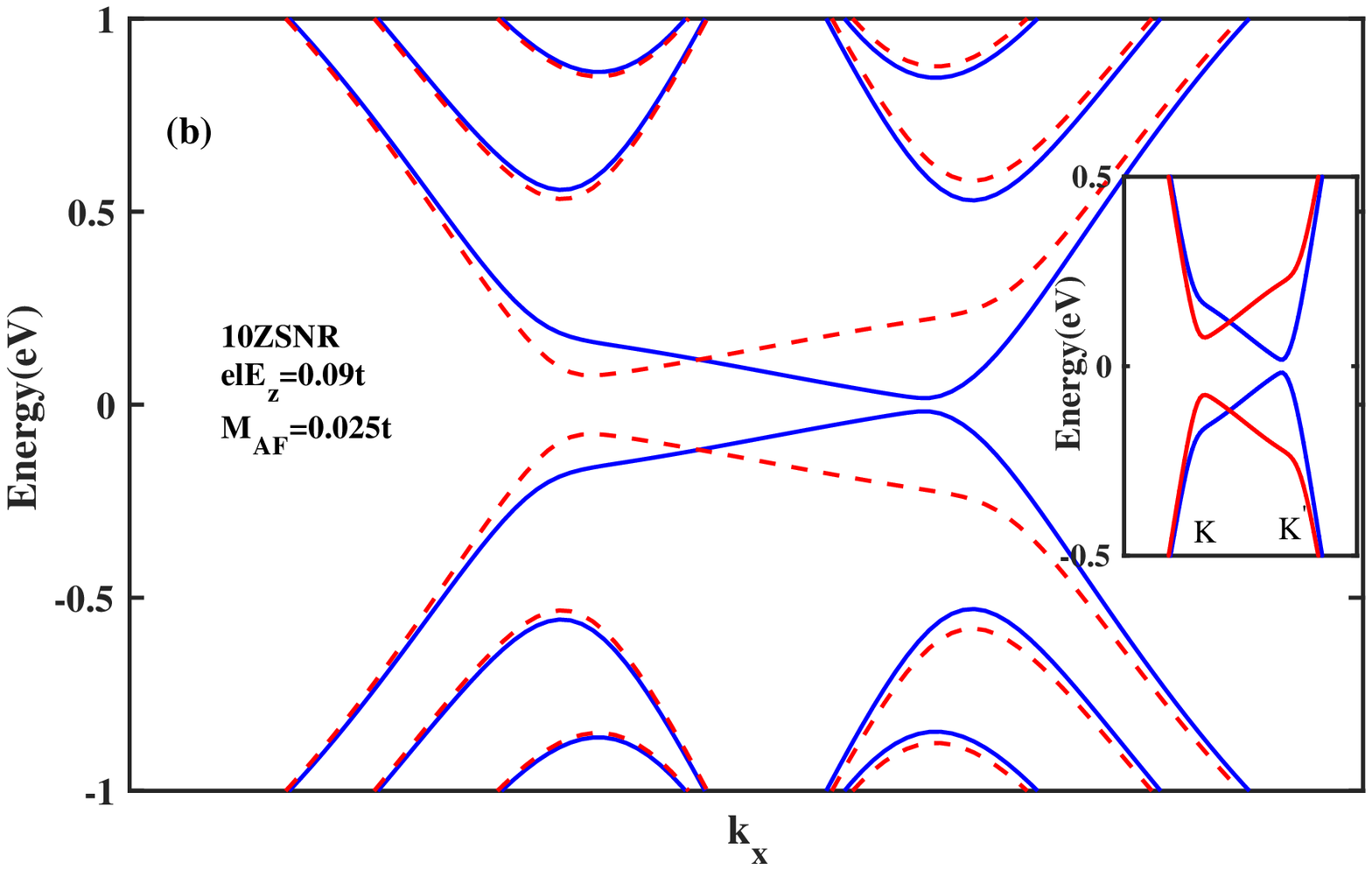}	
		\includegraphics[height=7cm]{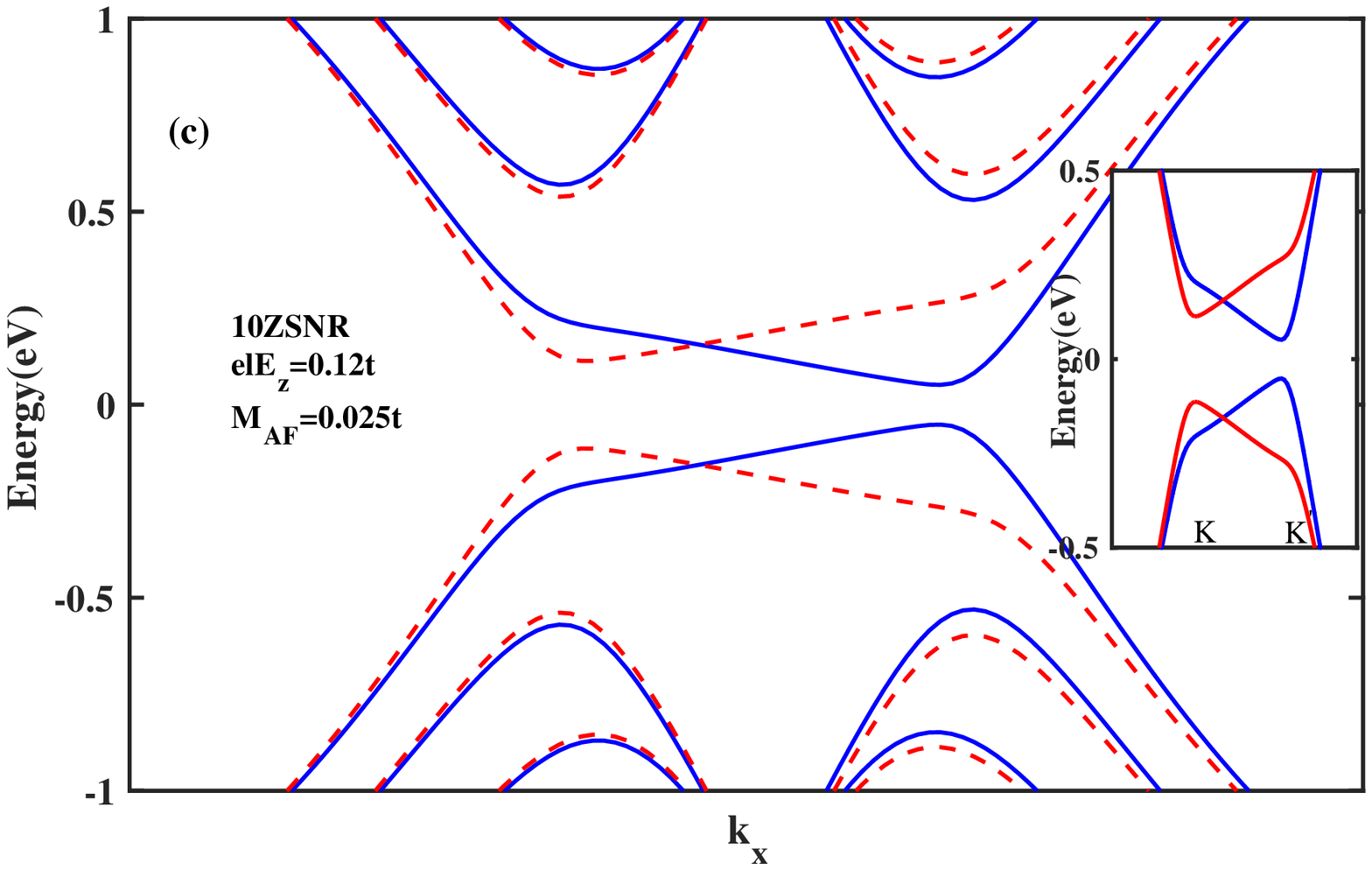}
	\end{center}
	\caption{The band structure of antiferromagnetic $ 10ZSNR $ subject to a perpendicular electric field: (a) with $ E_{z}=0 $, (b) with $ el E_{z}=0.09t $ and (c) with $ el E_{z}=0.12t $. Blue line denotes spin down and dashed red line denotes spin up.
	}\label{fig:1}
\end{figure*}

 In the first instance, spin transport properties of $ ZSNR $ with $ N=10 $ zigzag chains and $ 20 $ atoms in the unit cell are investigated. To this end, spin-resolved band structures of the antiferromagnetic $ ZSNR $ for various strengths of $ E_{z} $ is shown in Fig.~\ref{fig:1}. These results are similar to reports of previous study~\cite{tao2017all}. As can be seen in Fig.~\ref{fig:1}(a), in the absence of $ E_{z} $, twofold  spin degeneracy of the band structure  is observed. Applying the external electric field breaks inversion symmetry. The inversion symmetry breaking in combination with the large spin-orbit coupling lifts the band degeneracy in stanene. Also, it is found that switching the direction of the normal electric field will reverse the spin polarization of the band structure and hence,  sign of the spin polarized photocurrent. Moreover, finite energy band gap is emerged between spin-up and spin-down energy levels. (Figs.~\ref{fig:1} (b) and ~\ref{fig:1}(c)). As depicted in these figures, the magnitude of band gap is different for spin-up and spin-down energy levels. In the absence of antiferromagnetic term, the magnitude of band gap is equal for both spin states. Physically, due to  different effect of the antiferromagnetic term on the potential energies of spin states  in the on-site terms of the Hamiltonian, it is expected unequal displacement of spin-up and spin-down energy levels. Accordingly, the presence of the antiferromagnetic  exchange field leads to asymmetric  spin-dependent band gap and breaking of  time-reversal symmetry~\cite{luo2019antiferromagnetic}.
 
\begin{figure*}
	\begin{center}
		\includegraphics[height=7cm]{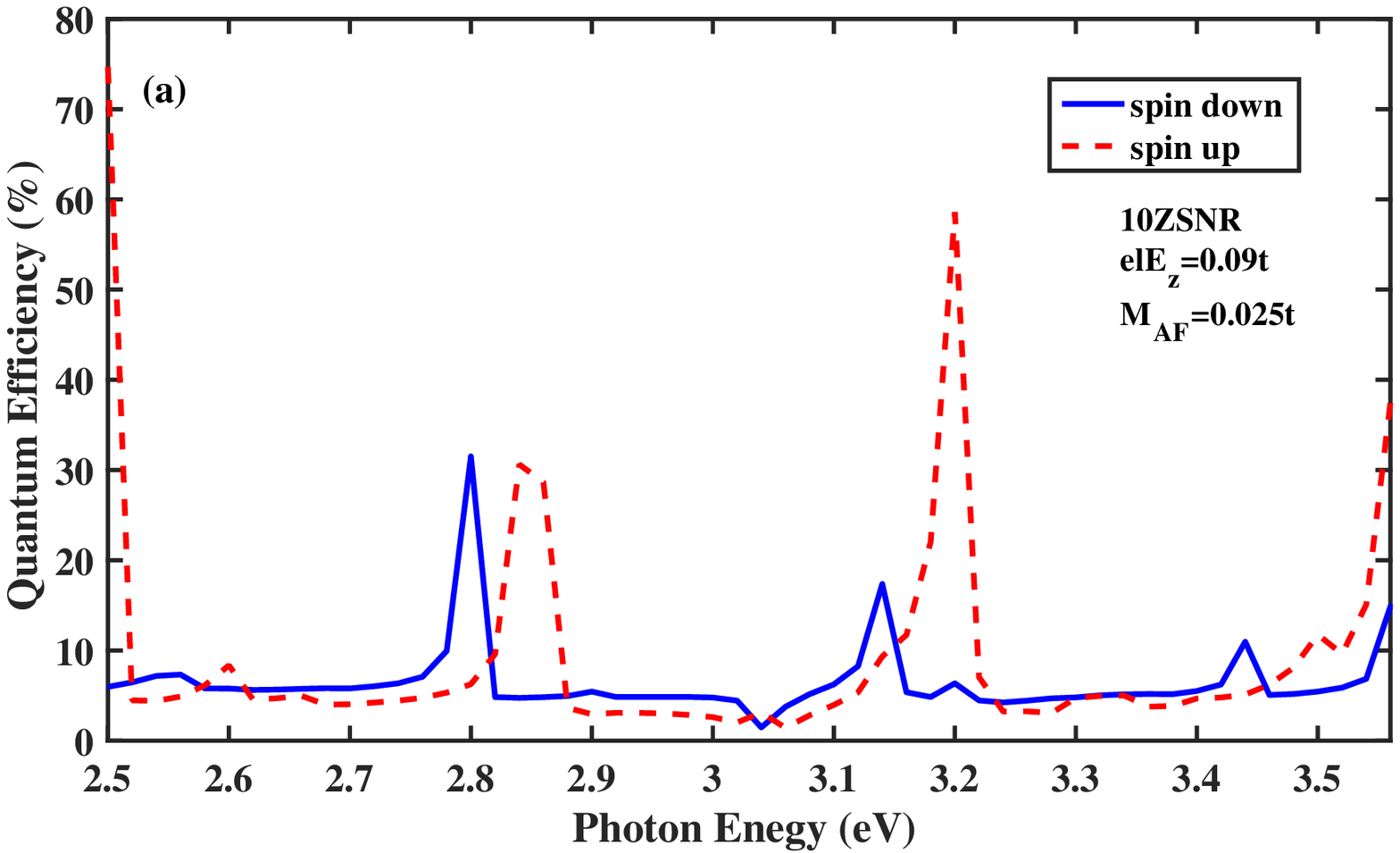}
		\includegraphics[height=7cm]{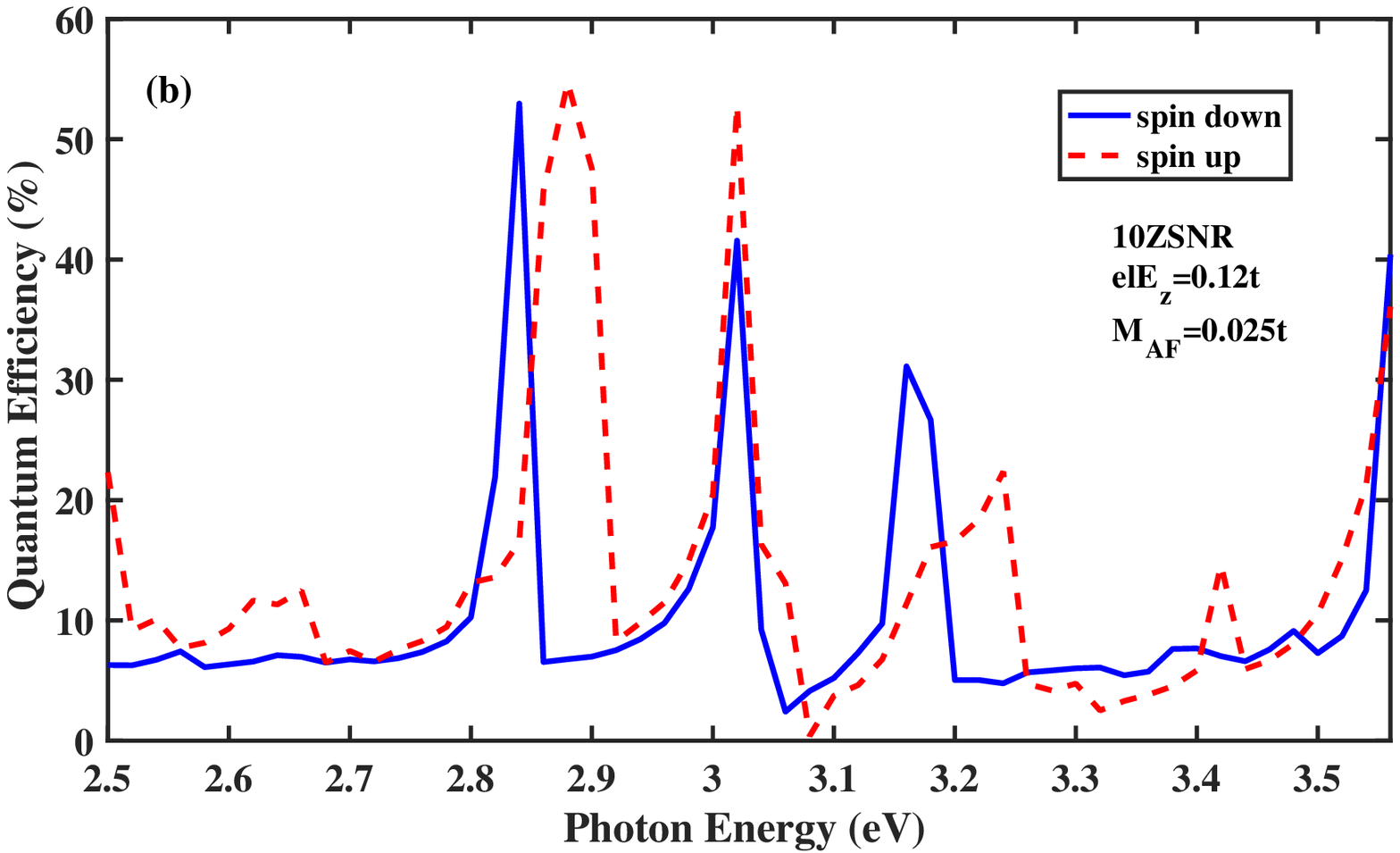}	
	\end{center}
	\caption{ Quantum efficiency as a function of the photon energy for the spin-photovoltaic device based on
		antiferromagnetic $10 ZSNR $ with $ M_{AF}=0.025t $ under the simultaneous effect of linear illumination with $ I_{w}=100\,\frac{kW}{cm^{2}} $  and (a)  $ el E_{z}=0.09t $, (b) $ el E_{z}=0.12t $.
	}\label{fig:2}
\end{figure*}

 The  spin-polarized quantum efficiency of $ 10ZSNR $ with $ M_{AF}=0.025t $ versus photon energy under the linearly polarized light and for different strengths of  $E_{z}$  is plotted in Fig.~\ref{fig:2}. As presented explicitly in the inset of Fig.~\ref{fig:1}(b),  the asymmetry of band gap energy at the $ K $ and the $K^{'}$ valleys results in  unequal  absorption of light which leads to spin population imbalance in these valleys.  It can be said  that these valleys in $ ZSNR $ are approximately equivalent to the valleys of infinite stanene sheet which are around the Dirac points~\cite{shahabi2020normal}. Also, spin splitting of the energy band structure  gives rise to  spin-dependent absorption when the linearly polarized illumination is shed normally on the top of central region. On the other hand, because of the relatively large enough spin  relaxation time in stanene~\cite{kurpas2019spin}, the electron preserves its spin in  this photo-excited phenomena. In fact, the spin-polarized carriers are excited from spin up(down) valence sub-bands to  spin up(down) conduction sub-bands by photons with an appropriate energy. Based on these discussions, different occupation numbers between the  $ K $  and the  $ K^{'} $ ($  K^{'}=-K $ ) valleys, which has been induced by the antiferromagnetic exchange field, leads to a unequal spin-polarized photocurrent for spin-up and spin down components. It should be mention that the normal electric field and the antiferromagnetic exchange field is applied to the scattering region.
 
 As can  be observed in Fig.2, the acceptable quantum efficiency is obtained for the photon energy within the range $ 2.5\:eV < E_{ph}  <3.56\:eV$. From the results represented in Fig.2, one can see considerable variations in spin optoelectronic properties of $ ZSNR $ in  the whole  allowable  photon energy range. In addition, position and magnitude of the most probable optical transitions are varied by different strengths of  $E_{z}$. It is obvious that the electric field changes  magnitude of the first optical absorption considerably.  
 
\begin{figure*}
	\begin{center}
		\includegraphics[height=9cm]{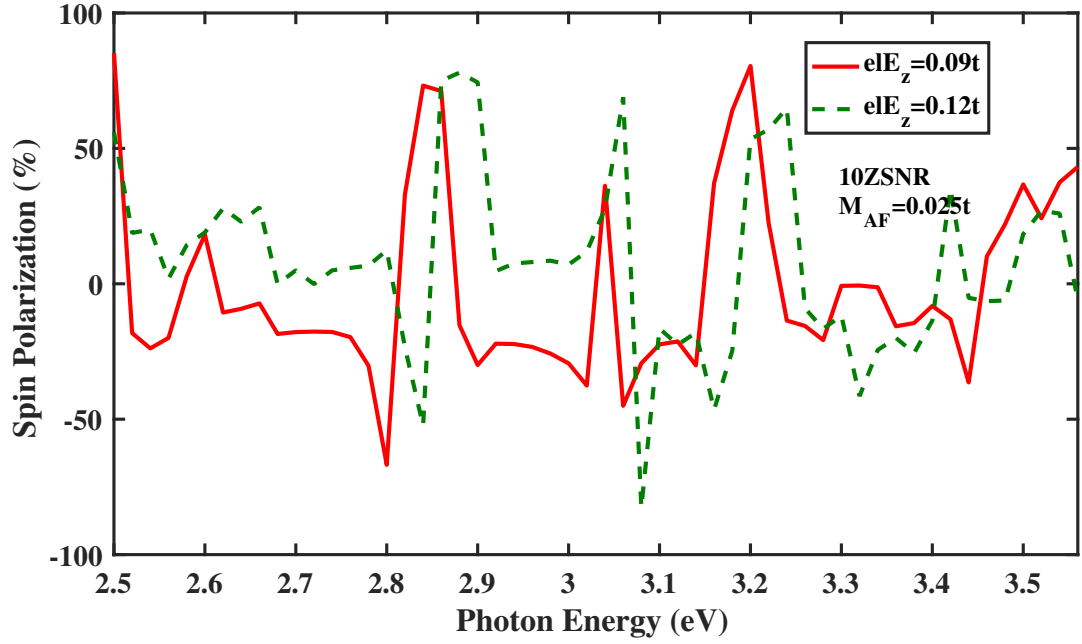}	
	\end{center}
	\caption{The optical spin polarization as a function of the photon energy for the spin-photovoltaic device based on antiferromagnetic $10 ZSNR $ with $ M_{AF}=0.025t $ for various $ el E_{z} $ strengths.
	}\label{fig:3.eps}
\end{figure*}

 For more clarification, optical spin polarizations of $ ZSNR $ for various strengths of $E_{z}$ is displayed in Fig.~\ref{fig:3.eps}. The spin polarization diagrams for different magnitudes  of the electric field have similar qualitative behavior, nearly.
 
 Moreover,  spin optoelectronic properties of other narrow ribbons with various widths has been studied. The  results reveal similar qualitative behavior but with differences in the magnitudes, positions of the highest peaks of quantum efficiency and acceptable  range of the photoresponsivity. These results demonstrate  the robustness of the obtained results to the ribbon size.\\
 
\begin{figure*}
	\begin{center}
		\includegraphics[height=7cm]{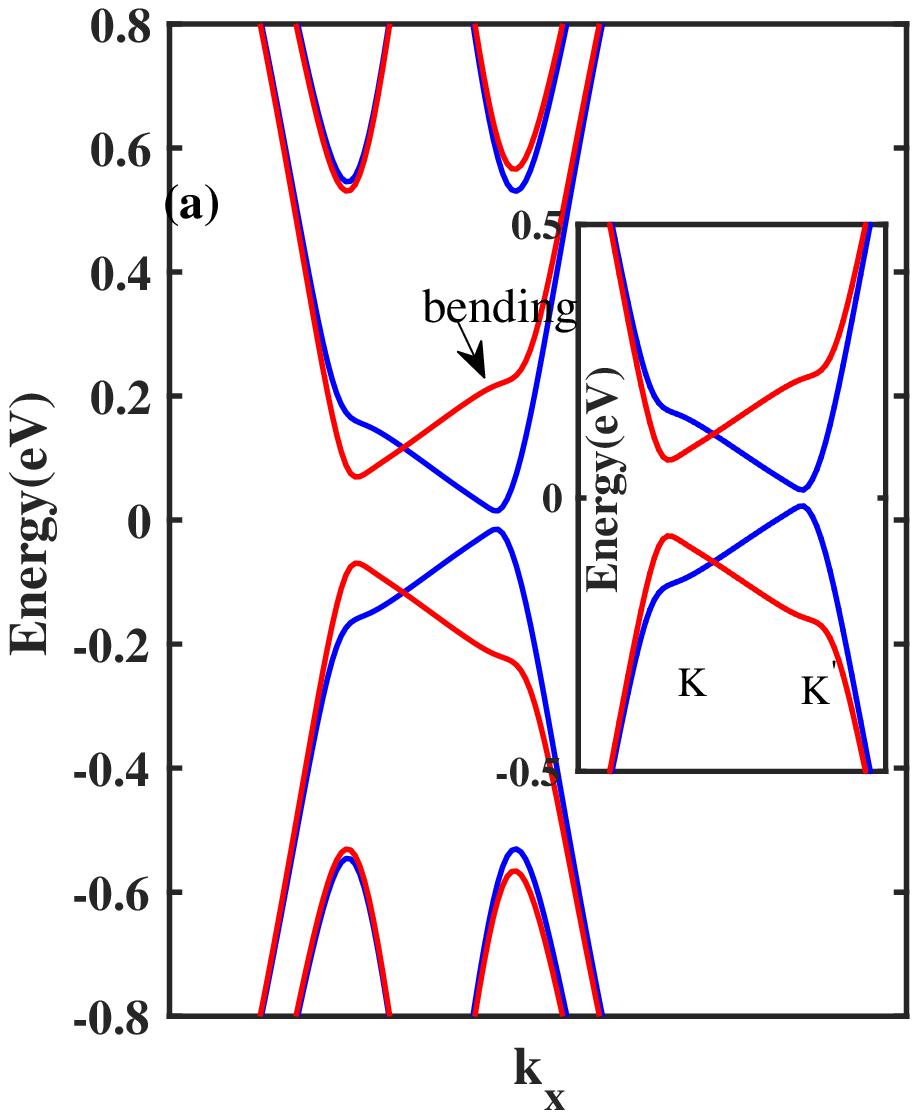}
		\includegraphics[height=7cm]{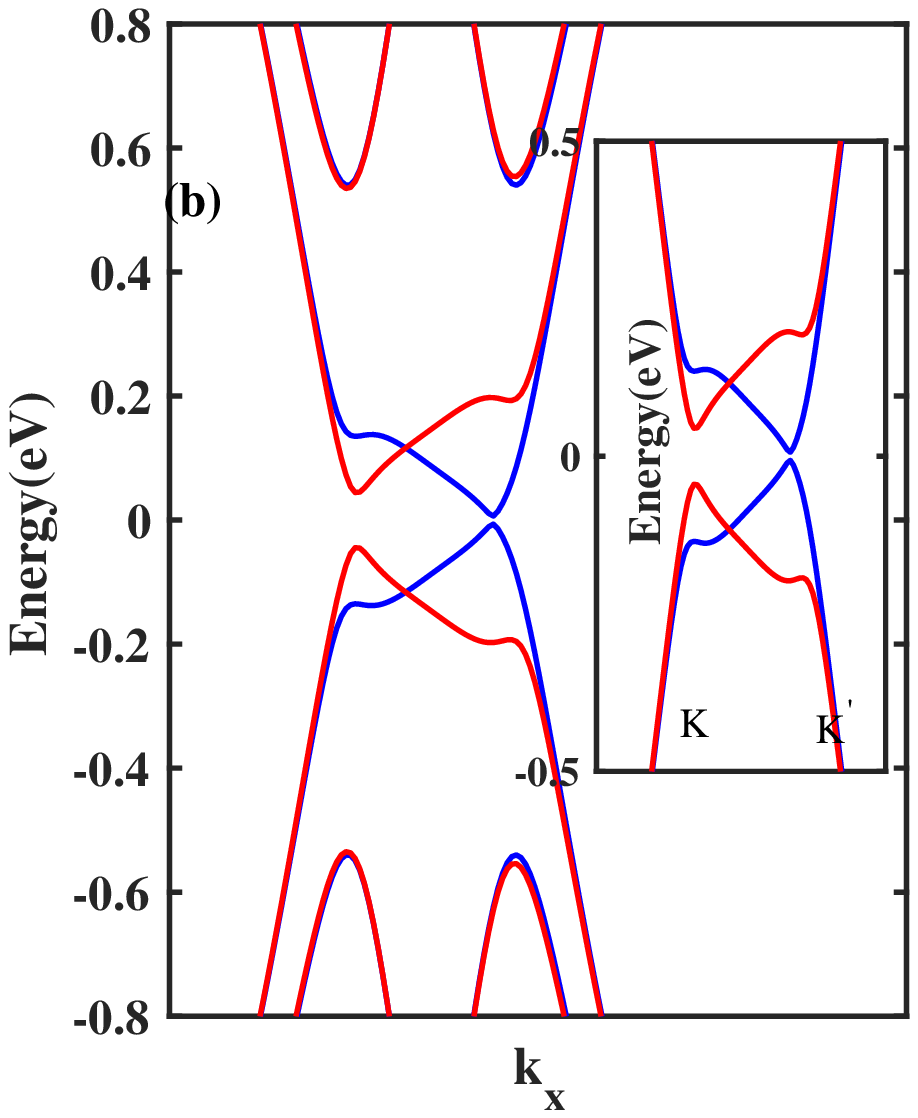}
		\includegraphics[height=7cm]{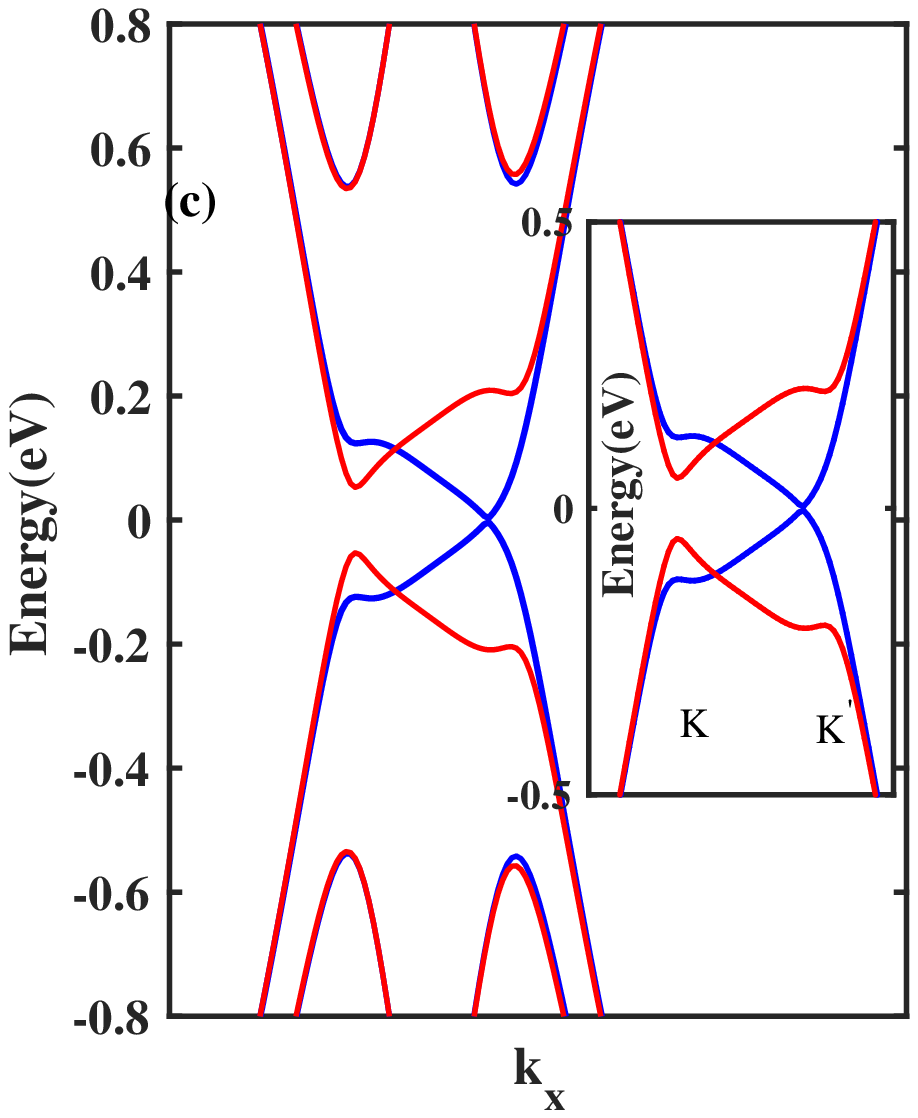}	
	\end{center}
	\caption{The band structure of antiferromagnetic $ 10ZSNR $ subjected to a edge electric field: (a) with $ M_{AF}=0.025t$ and applied electric field to $ N=8 $ zigzag chains, (b) $ M_{AF}=0.025t $ and applied electric field to $ N=4 $ zigzag chains, (c) $ M_{AF}=0.03t $ and applied electric field to $ N=4 $ zigzag chains. Blue line denotes spin down and red line denotes spin up. Here, $ el E_{z}=0.09t $.
	}\label{fig:4}
\end{figure*}

  {\label{sec:level2}Band bending}\\
  In this section the effect of band bending on the spin transport properties of single layer antiferromagnetic $ ZSNR $ is investigated. A previous study on the band structure and conductance of a zigzag silicene nanoribbon has shown that  band bending could be created  and controlled by the edge potentials. The bending near valleys can be realized by the edge states~\cite{lu2020band}.  In this paper, we proceed further by studying the benefits of both $ 2D $ buckled structure  and antiferromagnetic spintronics in order to harnessing  spin dependent transport in narrow $ ZSNRs $ in the presence of band bending. To this end, firstly,  the band structures of $ 10ZSNR$ in the presence of  edge electric field is analyzed (Fig.~\ref{fig:4}).  In Fig.~\ref{fig:4}(a), the edge field is applied to $ N=8 $ zigzag chains, such that two chains which are located at the center of ribbon are not affected by $E_{z}$. Also, in Fig.~\ref{fig:4}(b), the edge potential is applied to $ N=4 $ zigzag chains and six chains  at the center of ribbon are not affected by $E_{z}$. It should be noted that applied edge fields at the two edges of ribbon are symmetric, that is, $ \vec{E_{z1}}= \vec{E_{z2}}=\vec{E_{z}} \vec{z}$. Owing to the buckled structure of stanene,  the edge potentials could significantly impact on the edge states and the band structure. Fig.~\ref{fig:4}(a) and Fig.~\ref{fig:4}(b) indicate that with the narrowing of edge field, the bending enhances gradually, which is in good agreement with previous report~\cite{lu2020band}. Also, in Fig.~\ref{fig:4}(c) electronic band structure of $ 10ZSNR $ with narrow edge field  and $ M_{AF}=0.03t $ is displayed. By increasing the magnitude of antiferromagnetic exchange field from $ M_{AF}=0.025t $ to $ M_{AF}=0.03t $  in combination with applied edge field to $ N=4 $ chains, the half-metal behavior is revealed  at the band structure as shown in Fig.~\ref{fig:4}(c). For spin-down electrons, the gap is closed  and  metallic behavior is observed, while spin-up carries display still semiconductor properties.
  
\begin{figure*}
	\begin{center}
		\includegraphics[height=7cm]{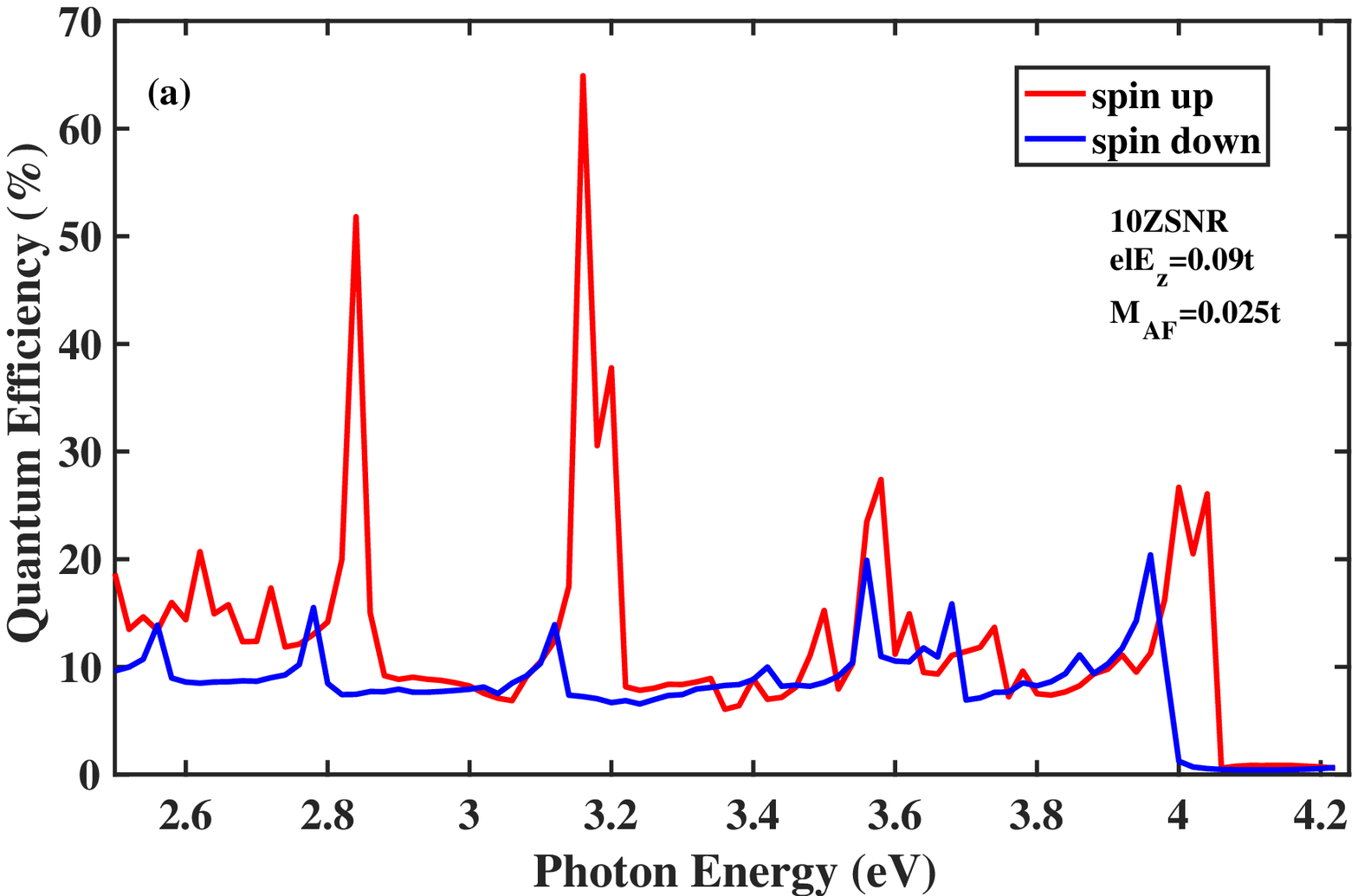}
		\includegraphics[height=7cm]{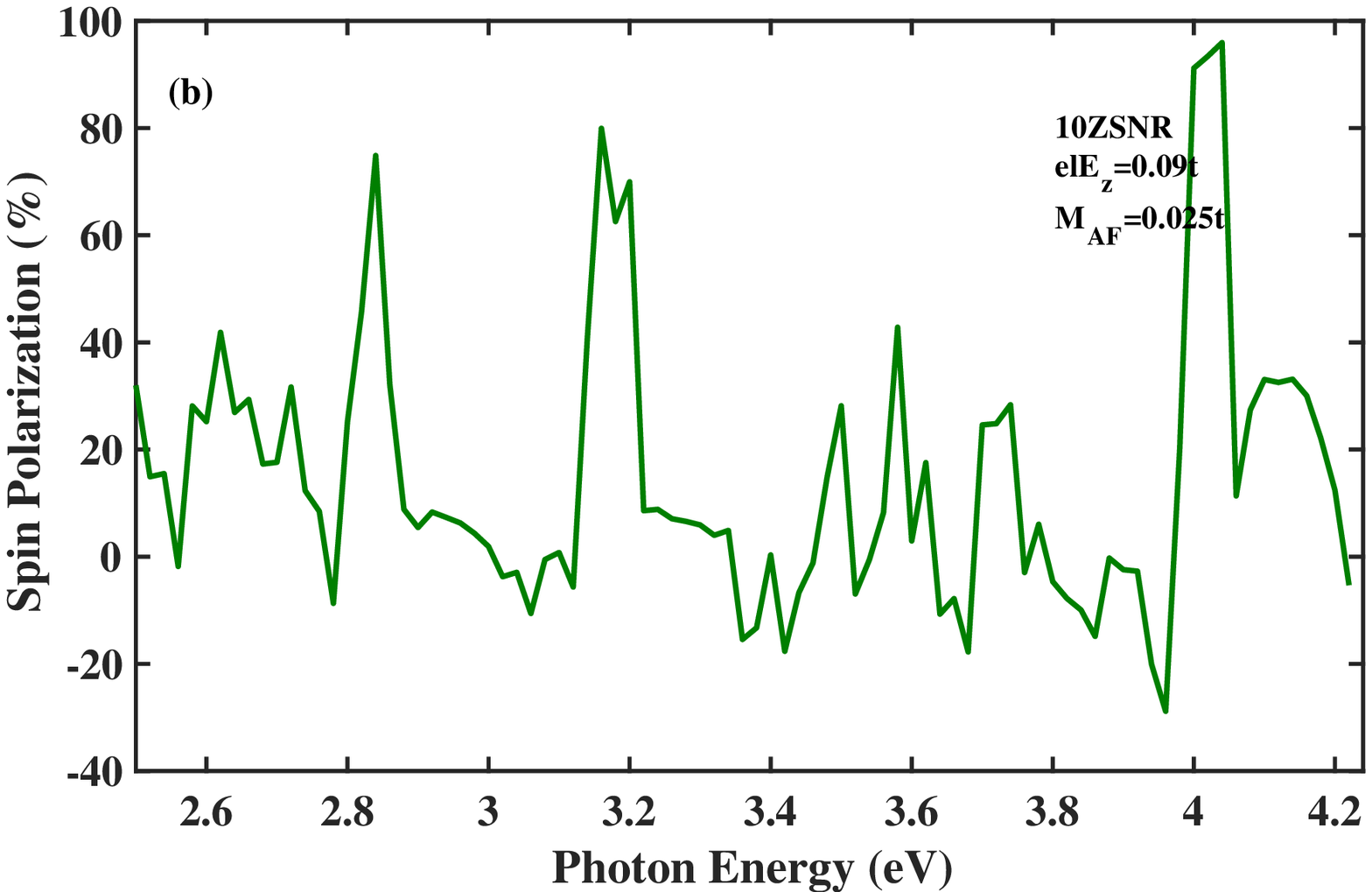}	
	\end{center}
	\caption{ (a) Quantum efficiency and (b) Spin polarization as a function of the photon energy for the spin-photovoltaic device based on antiferromagnetic	$10 ZSNR $  under the simultaneous effect of linear illumination with
		 $ I_{w}=100\,\frac{kW}{cm^{2}} $ and  applied edge field to $ N=8 $ chains. Other parameters are: $ M_{AF}=0.025t $ and  $ el E_{z}=0.09t $.
	}\label{fig:5}
\end{figure*}

\begin{figure*}
	\begin{center}
		\includegraphics[height=7cm]{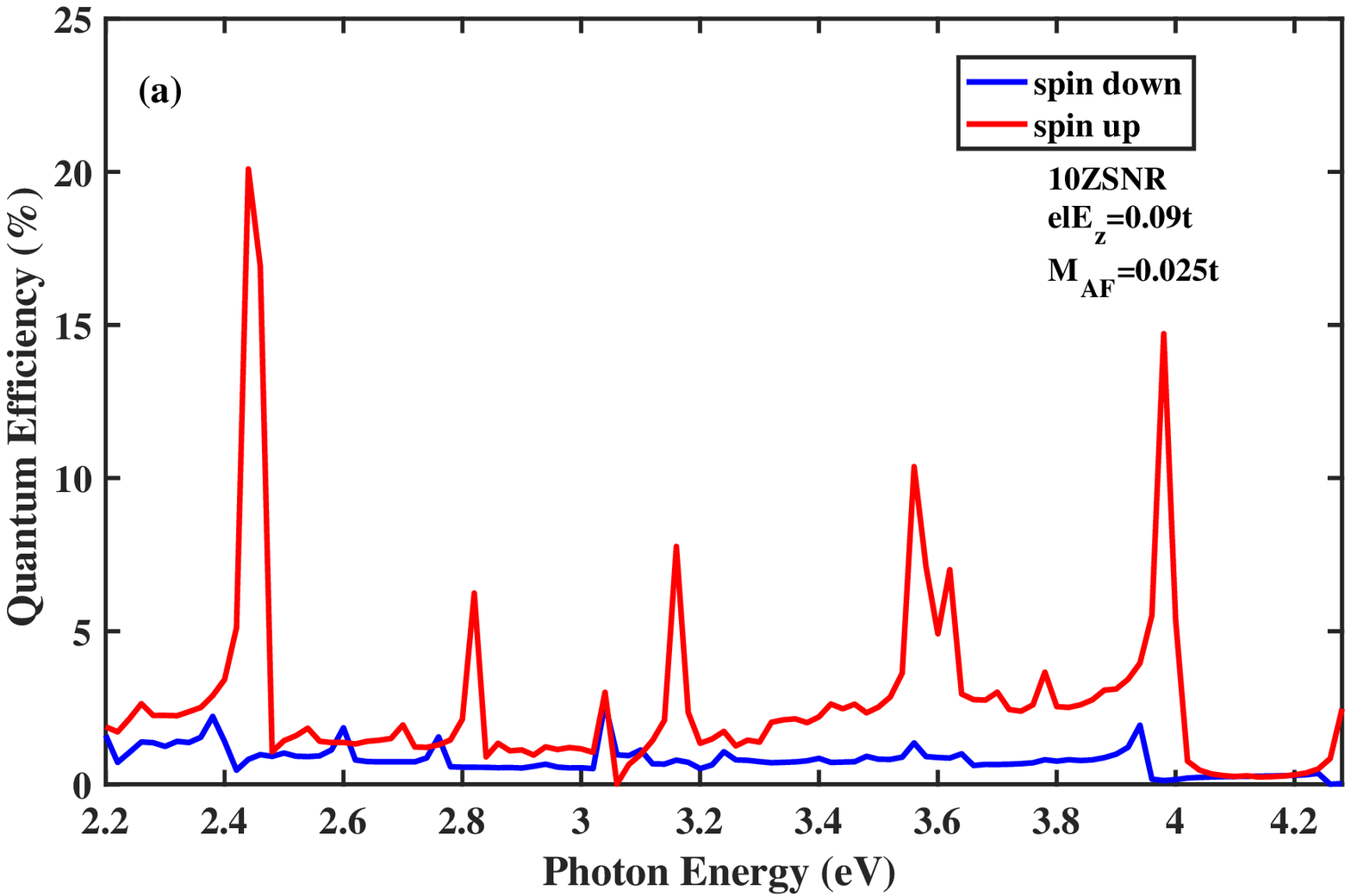}
		\includegraphics[height=7cm]{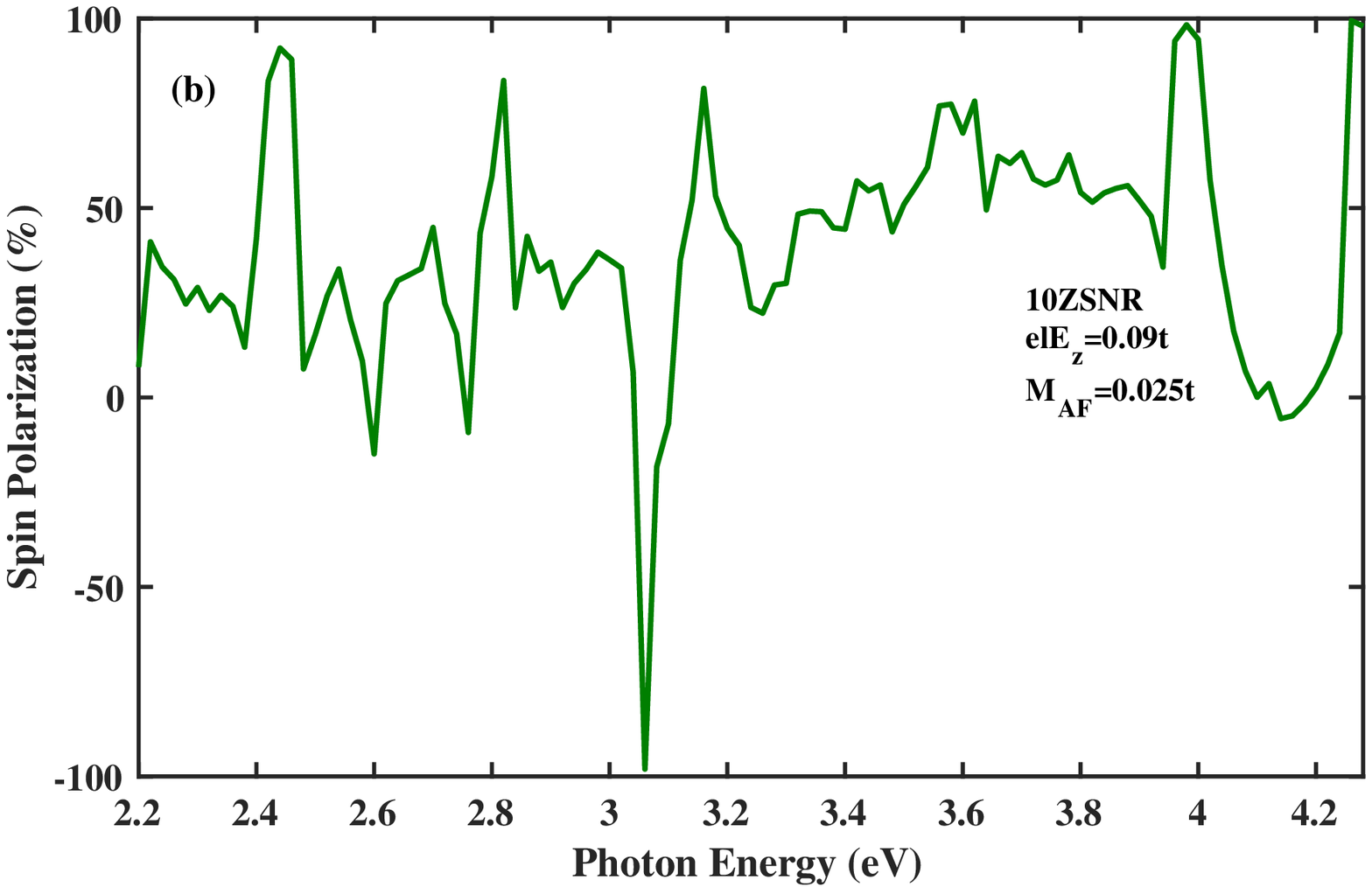}	
	\end{center}
	\caption{ (a) Quantum efficiency and (b) Spin polarization as a function of the photon energy for the spin-photovoltaic device based on antiferromagnetic	$10 ZSNR $  under the simultaneous effect of linear illumination with
		$ I_{w}=100\,\frac{kW}{cm^{2}} $ and  applied edge field to $ N=4 $ chains. Other parameters are: $ M_{AF}=0.025t $ and  $ el E_{z}=0.09t $.
	}\label{fig:6}
\end{figure*}

\begin{figure*}
	\begin{center}
		\includegraphics[height=7cm]{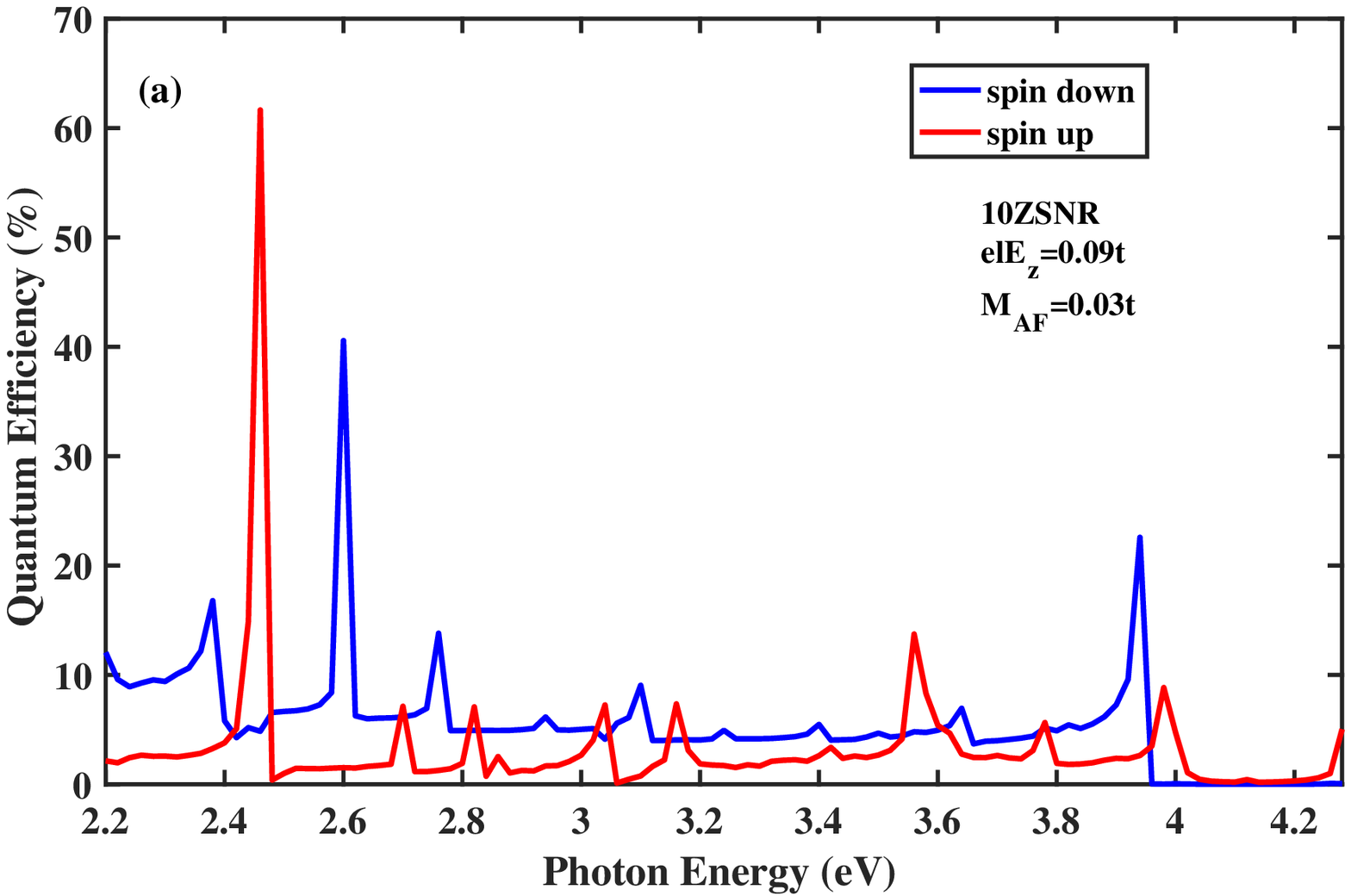}
		\includegraphics[height=7cm]{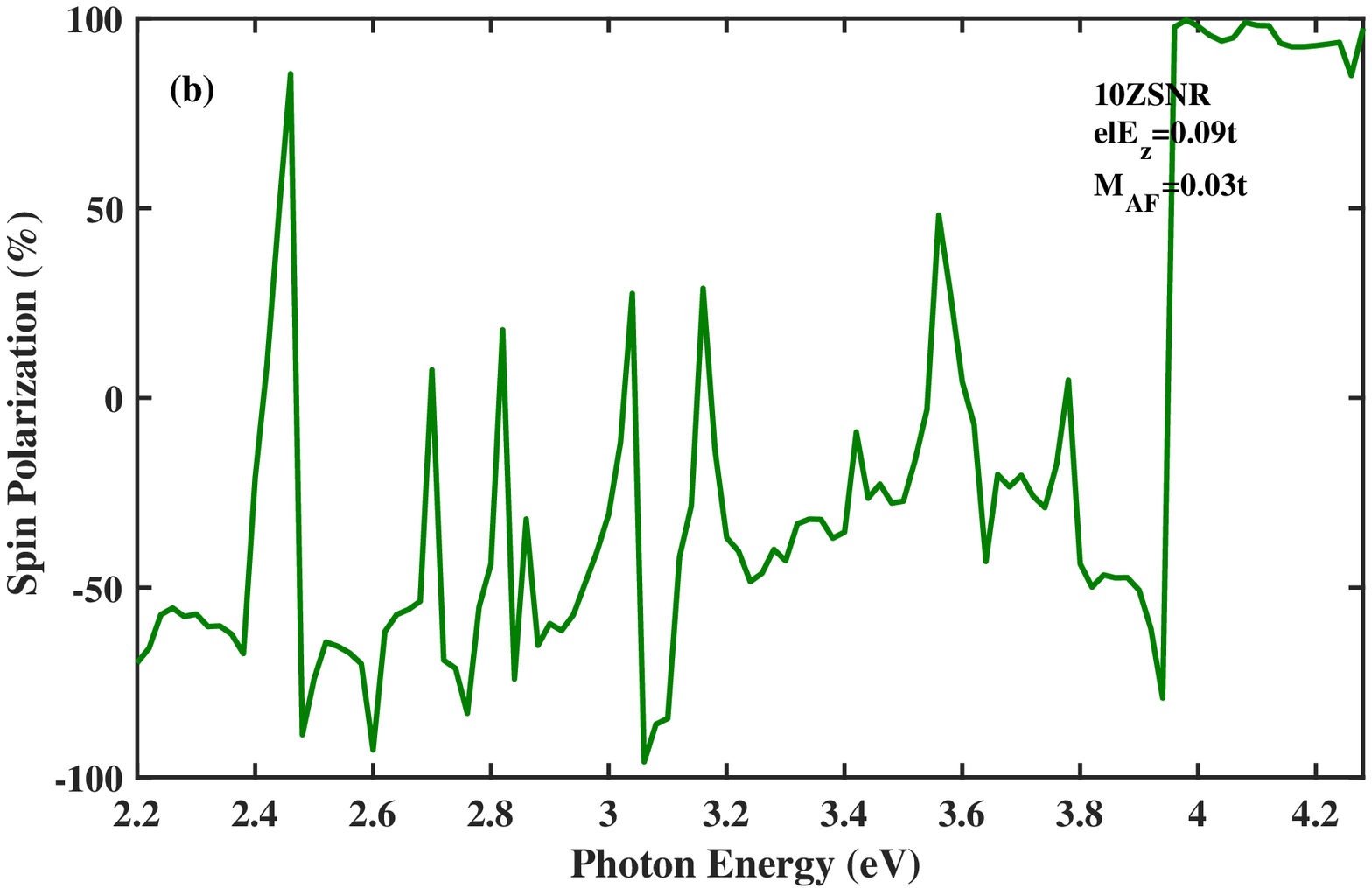}	
	\end{center}
	\caption{ (a) Quantum efficiency and (b) Spin polarization as a function of the photon energy for the spin-photovoltaic device based on antiferromagnetic	$10 ZSNR $  under the simultaneous effect of linear illumination with
		$ I_{w}=100\,\frac{kW}{cm^{2}} $ and  applied edge field to $ N=4 $ chains. Other parameters are: $ M_{AF}=0.03t $ and  $ el E_{z}=0.09t $.
	}\label{fig:7}
\end{figure*}

   Quantum efficiency as a function of  photon energy for the spin-photovoltaic device based on	antiferromagnetic $10 ZSNR $ under applied edge field to $ N=8 $ chains is presented in Fig.~\ref{fig:5}. Note that, here,  $ el E_{z}=0.09t $ and  $ M_{AF}=0.025t $.  In comparison with Fig.~\ref{fig:2}(a), it is clear that , in general, quantum efficiency is enhanced for both spin states in the presence of edge potential. Note that, in Fig.~\ref{fig:2}, the electric field is  applied to  the whole area of the central region. Furthermore, acceptable range of the photoresponsivity is broaden under edge potential.  In this case, the maximum magnitude of the spin polarization is obtained $ 95.94\:\% $. at $ E_{ph}=4.04\: eV $.  In Figs.~\ref{fig:6}(a) and ~\ref{fig:6}(b), quantum efficiency and spin polarization  of $10 ZSNR $ under applied edge field to $ N=4 $ chains are computed, respectively. Other parameters are: $ el E_{z}=0.09t $ and  $ M_{AF}=0.025t $. By comparing Fig.~\ref{fig:6}(a) with Fig.~\ref{fig:5}(a), it is clear that allowable range of the photoresponsivity is broaden with the narrowing of the edge field. Additionally, the identical qualitative manner of the quantum efficiency is realized, although  position and  height of the  spin-dependent optical transition lines are different, which comes from the difference in the spin-resolved band structure  for spin-up and  spin-down states. By evaluating of the spin polarization diagrams in Fig.~\ref{fig:6}(b)  and  Fig.~\ref{fig:5}(b), one can see that  spin polarization is improved with the narrowing of the edge potential. The highest spin polarization peak of $ 99.43\:\% $ is appeared at $ E_{ph}=4.26\: eV $. The effect of the antiferromagnetic exchange field strength increasing on spin optoelectronic behavior is investigated in Fig.~\ref{fig:7}. In this figure, the edge potential is applied to the $ N=4 $ zigzag chains with $ el E_{z}=0.09t $ and here $ M_{AF}=0.03t $. Obviously in Fig.~\ref{fig:7}(a), quantum efficiency of spin-down carries is higher than quantum efficiency of spin-up carries  for a broad range of photon energies which is in contrast to the observed overall trend for quantum efficiency in the presence of the edge field (Fig.~\ref{fig:5}(a) and Fig.~\ref{fig:6}(a)). This distinguished behavior can be attributed to the half-metallic feature of spin-down component. Moreover, as can be observed in Fig.~\ref{fig:7}(b), the perfect $ (100\%) $ optical spin polarization is obtained at $ E_{ph}=3.98\: eV $.\\
   
     {\label{sec:level2}Strain}\\ 
     In the crystal structure of a unstrained single-layer stanene and  equilibrium condition, each atom of the upper sublattice is connected to its neighboring atoms  in the lower sublattice through  three bond vectors: $R_{1}^{0}=\frac{a}{\sqrt{3}}(\frac{\sqrt{3}}{2},\frac{1}{2},\cot \varphi)  $, $R_{2}^{0}=\frac{a}{\sqrt{3}}(-\frac{\sqrt{3}}{2},\frac{1}{2},\cot \varphi) $ and $R_{3}^{0}=\frac{a}{\sqrt{3}}(0,-1,\cot \varphi)$, where $ a=4.7 \hspace{2mm}\AA{} $ is lattice constant of stanene and $ \varphi=107.1^{0} $.  The presence of an external tension $ \textbf{T} $, which is defined as $ \textbf{T}=T\cos \theta \hspace{2mm} \hat{e_{x}}+ T\sin \theta\hspace{2mm} \hat{e_{y}}$, leads to the transformation of these vectors. Longitudinal expansion of stanene layer arising from tension is determined by $ \epsilon=(\acute{a}-a)/a$. The tight-binding approximation is a standard approach to describe the electronic properties of nanodevices. One of the important benefits of this approximation is that  one can take into account  the effect of strain only by modifying the tight-binding parameters. By applying a uniaxial strain in stanene, equilibrium distance, $ R_{0}=a/(\sqrt{3}\sin \varphi)$, is distorted. Thereby, the hopping energy is changed. Calculations demonstrate that under the effect of strain, hopping coefficients are modified as follows~\cite{farokhnezhad2017strain}:
     \begin{eqnarray}\label{algebra4}
     &&t_{n}(R_{n})=t(1-\beta\,\frac{\delta\,R_{n}}{R_{0}}),\quad n=1,2,3
     \end{eqnarray}
     where $ \beta $ is Gr\"{u}neisen parameter and in our simulation we take $ \beta=1.95 $~\cite{yan2015tuning, huang2015phonon}.
     
     To study the effects of a uniaxial strain on the performance of  proposed spin-optoelectronic device  based on antiferromagnetic ZSNRs,  the  spin-dependent  band structure and the quantum  efficiency under  various strains is inspected. Experimentally, in 2D materials, local strains are generated by  depositing them on a prestretched elastomeric~\cite{castellanos2013local} or rough substrates~\cite{shin2016indirect}.
     
\begin{figure*}
	\begin{center}
		\includegraphics[height=7cm]{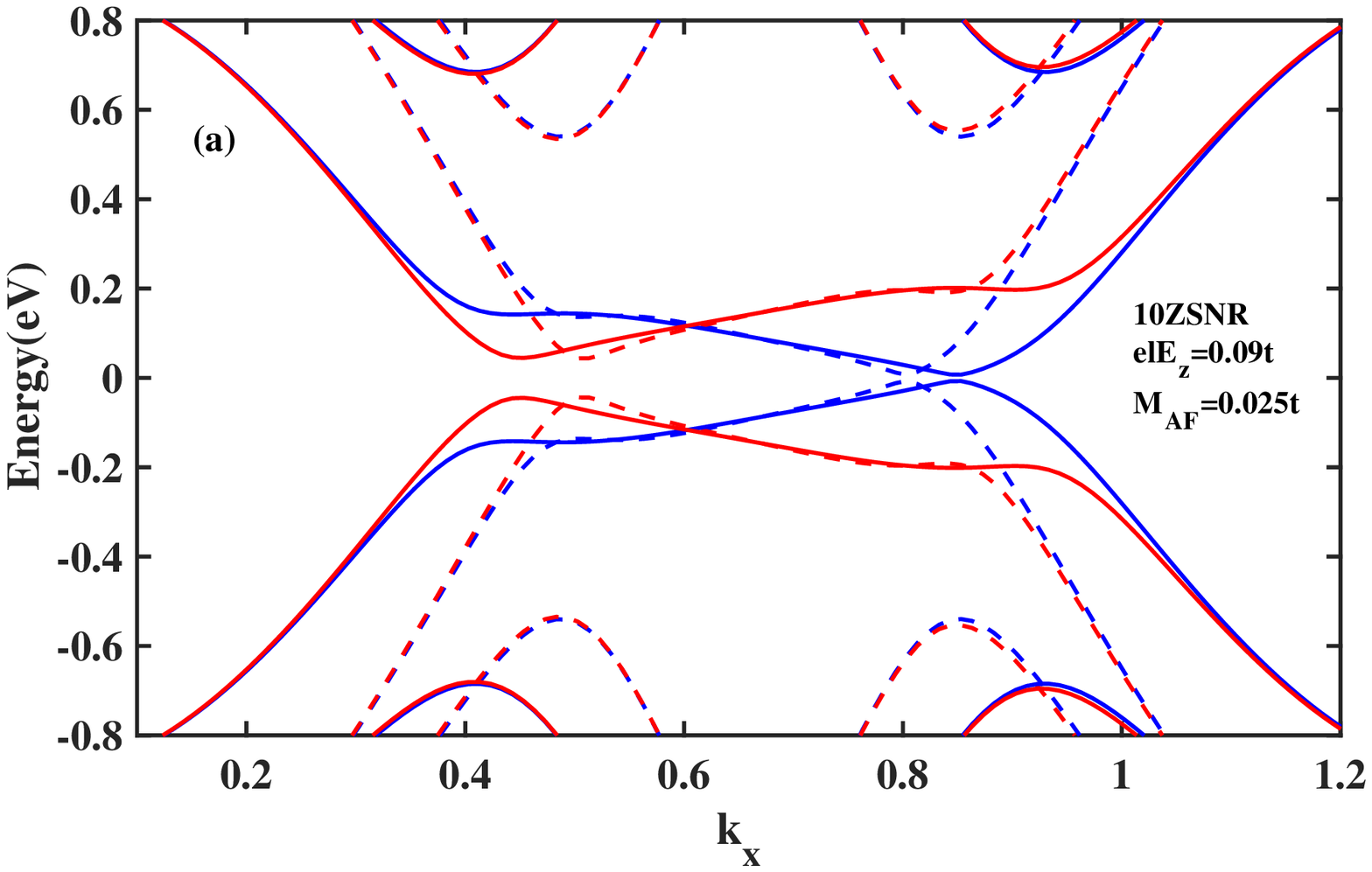}
		\includegraphics[height=7cm]{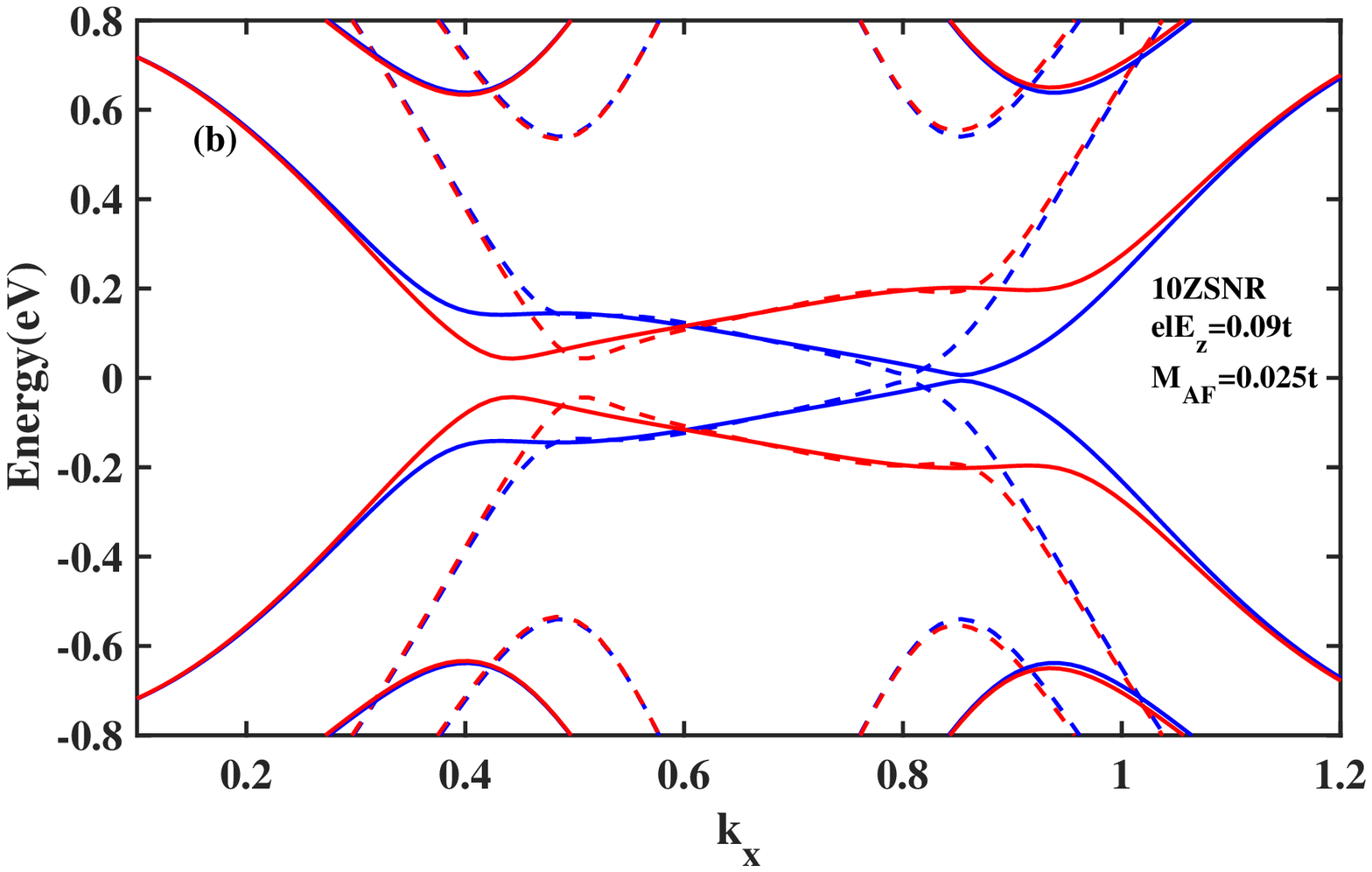}
		
	\end{center}
	\caption{ The band structure of antiferromagnetic $ 10ZSNR $ subjected to the combining effect of strain and narrow edge potential
		with applied edge field to $ N=4 $ chains. (a) $ \epsilon=0.2 $ and $ \theta=0^{\circ} $ and (b) $ \epsilon=-0.2 $ and $ \theta=90^{\circ} $. Blue line denotes spin down and red line denotes spin up.
		 Also, dashed blue line denotes spin down and dashed red line denotes spin up in the absence of strain. 
	}\label{fig:8}
\end{figure*}

     The results mentioned in this sub-section  are for the case in which spin-optoelectronic device is based on antiferromagnetic $ 10ZSNR $ and  edge potential is applied to $ N=4 $ zigzag chains where six chains  at the center of ribbon are not affected by $E_{z}$. As depicted in Fig.~\ref{fig:8}, applying the uniaxial strain leads to a shift in the spin-resolved sub-bands around the Fermi energy and it therefore results in change of population of the states. As a consequence, by employing  strain, one can monitor  accumulation of spin-polarized carriers in the scattering region. In this case,  the variation of the kinetic  energy induced by strain in the hopping
     term of the Hamiltonian is included.  This may lead to the alteration of magnitude and also the number of  peaks corresponding to spin-up  and spin-down components. 
     
\begin{figure*}
	\begin{center}
				\includegraphics[height=7cm]{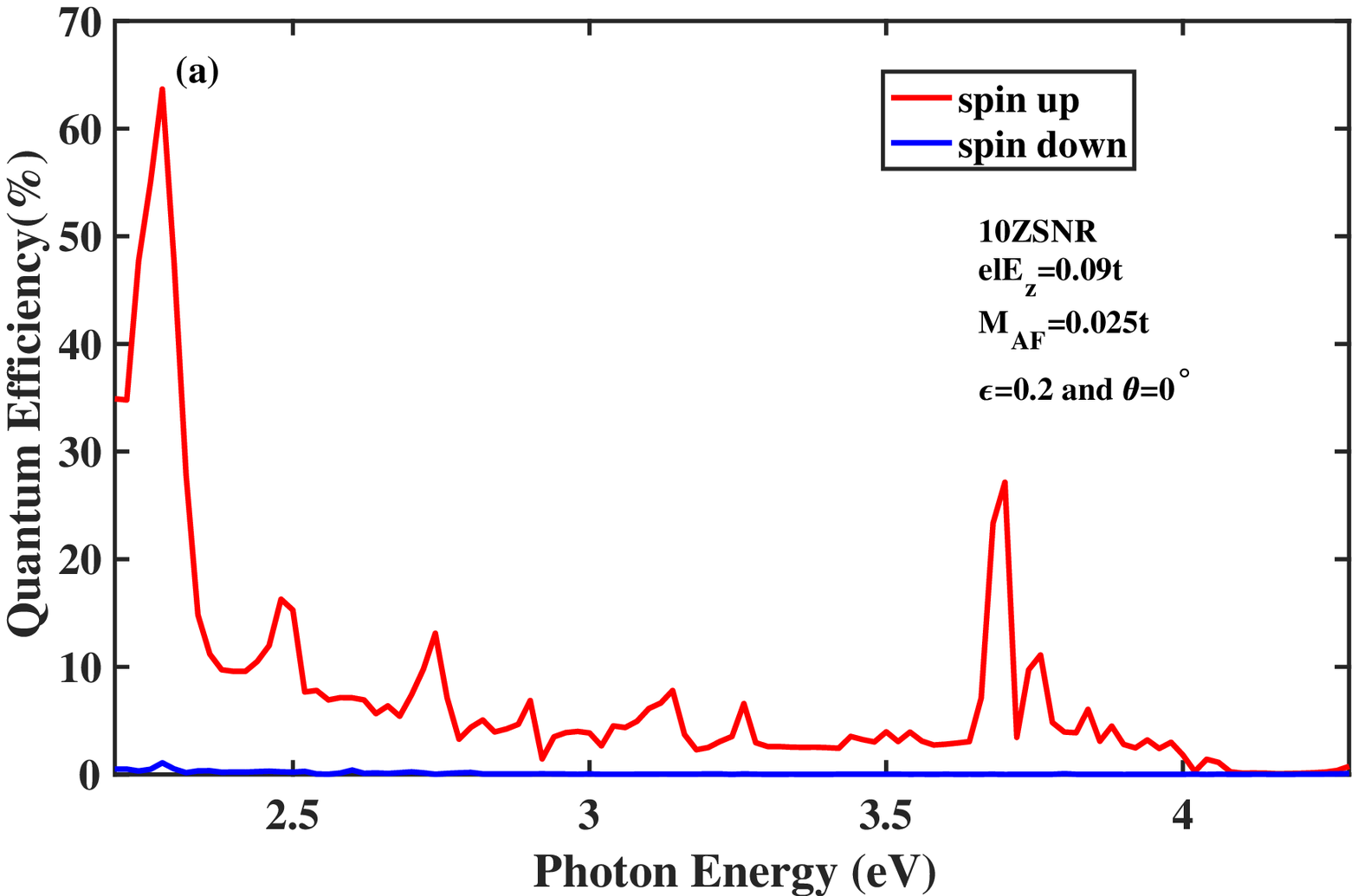}
				\includegraphics[height=7cm]{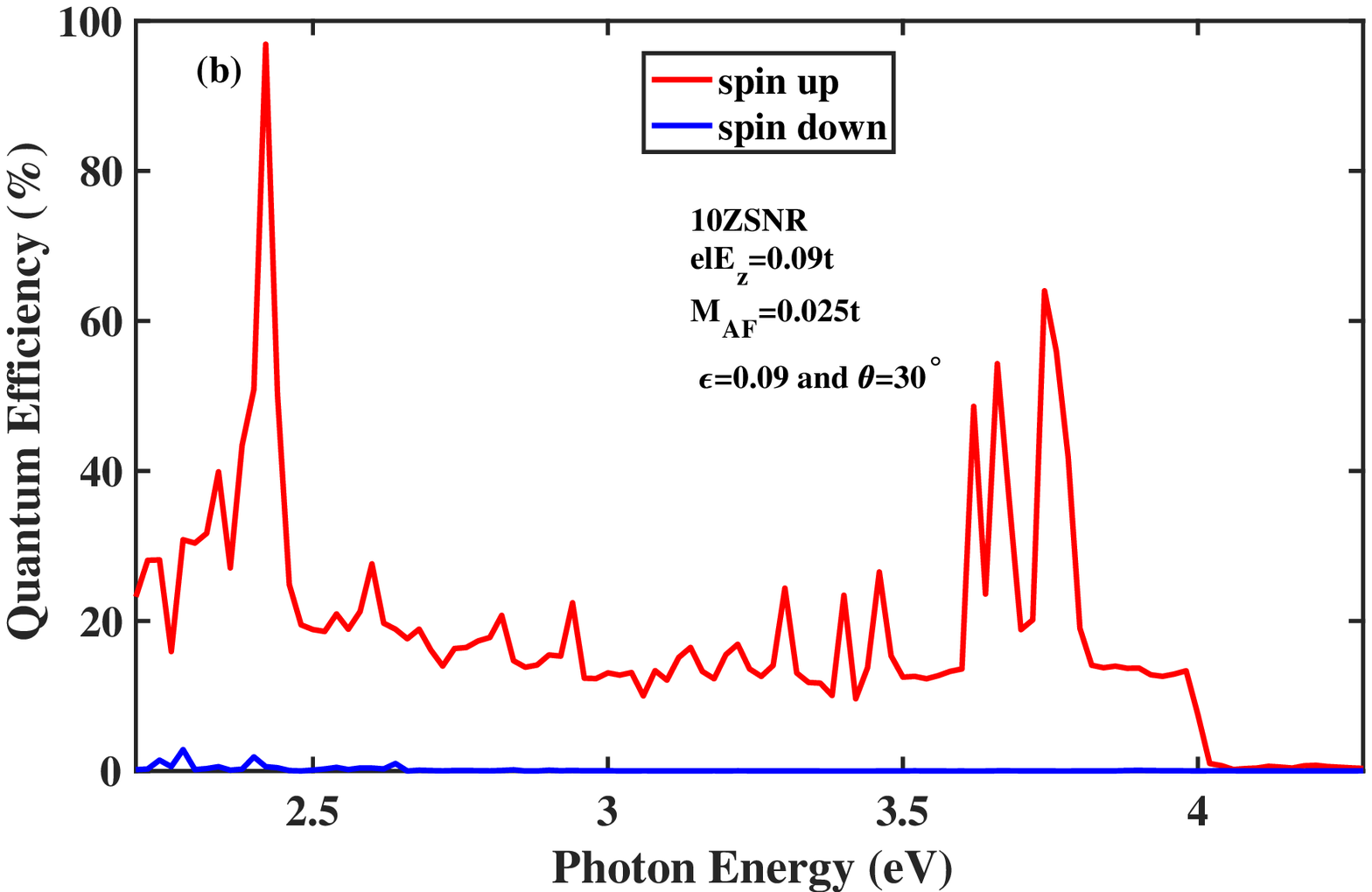}
				\includegraphics[height=7cm]{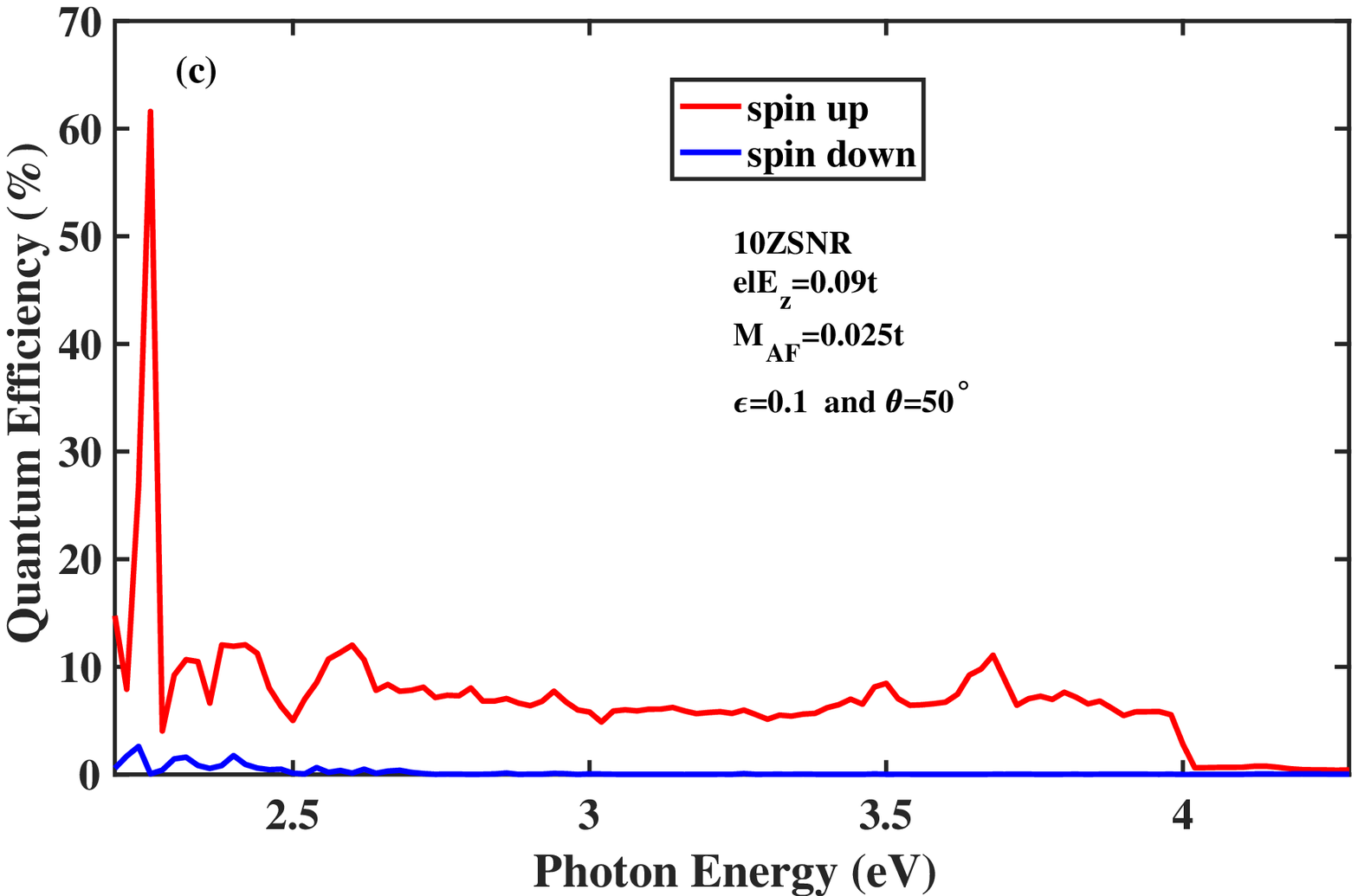}

	\end{center}
	\caption{  Quantum Efficiency as a function of the photon energy for the spin-photovoltaic device based on antiferromagnetic $10 ZSNR $  under the combining effect of strain and narrow edge potential
		with applied edge field to $ N=4 $ chains. Other parameters are: $ I_{w}=100\,\frac{kW}{cm^{2}} $, $ M_{AF}=0.025t $ and  $ el E_{z}=0.09t $..
	}\label{fig:9}
\end{figure*}

     Fig.~\ref{fig:9} presents the numerical results for the spin-dependent quantum efficiency of $ 10ZSNR $ for different applied tensions. As can be seen  in this figure, different tensions reveal spin-optoelectronic  features  in an energy range between  $ 2.2\:eV < E_{ph}  <4.28\:eV $. In addition,  the similar qualitative behavior for the spin-dependent  quantum efficiency diagram is obtained in the presence of  different strains. Nevertheless,  due to the variation in spin-resolved energy levels and the existence  of various band gaps for spin-up and spin-down states for different strains,  location and  height of the spin-dependent optical transition  peaks  are varied. Moreover, by comparing Fig.~\ref{fig:9} with Fig.~\ref{fig:6}(a)  it can be understood  that in the presence of strain the magnitude of quantum efficiency is  decreased strongly for spin down component and moderately increased for spin up component. Accordingly, it is expected  that applying strain leads to spin-filtering effect  in the spin optoelectronic device.
     
\begin{figure*}
	\begin{center}
		\includegraphics[height=9cm]{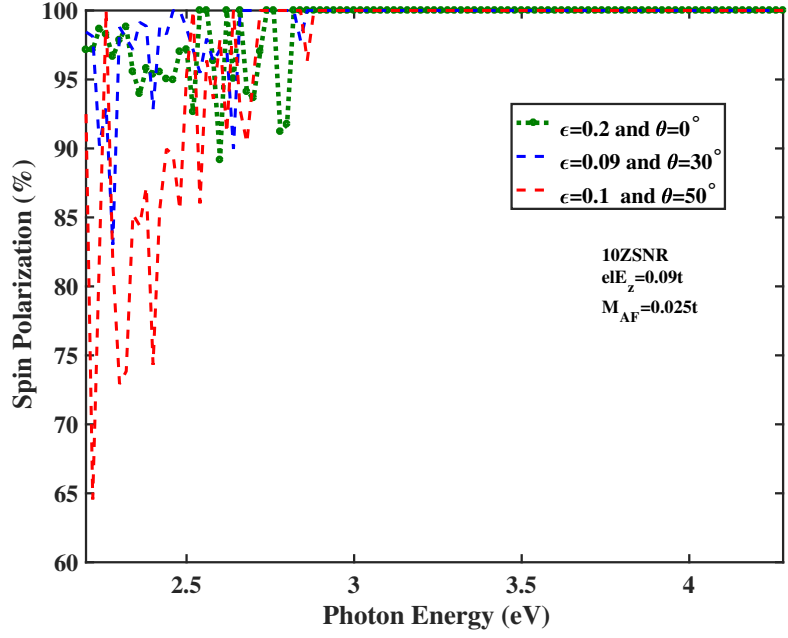}		
		
	\end{center}
	\caption{ Optical spin polarization as a function of the photon energy for the spin-photovoltaic device based on antiferromagnetic $10 ZSNR $  under the combining effect of strain and narrow edge potential
		with applied edge field to $ N=4 $ chains. Other parameters are: $ I_{w}=100\,\frac{kW}{cm^{2}} $, $ M_{AF}=0.025t $ and  $ el E_{z}=0.09t $.
	}\label{fig:10}
\end{figure*}

      In order to get an exact explanation of  spin-filtering effect in the presence of strain,  the optical spin polarization of  ZSNR  as function of  the photon energy is shown  in Fig.~\ref{fig:10}  for different strengths of the  strain. As mentioned earlier, the spin splitting  of the electronic band structures which is attributed to the simultaneous effect  of vertical  electric field and large spin orbit coupling  leads to  appearance of the spin polarization and particularly, a fully spin-polarized photocurrent for only one spin state. Furthermore, the spin-filtering photoresponsivity which is induced under the effect of strain is occurred  for the wide photon energy range from $ 2.84\:eV $ to $ 4.28\:eV $, approximately. It can be said that strain improves the optical spin-filtering property  of  stanene lattice such that  the full spin polarization range of the  photon energy  is broaden considerably. 
      
     {Conclusion} 
            
      In summary, spin-polarized photocurrent in the antiferromagnetic single layer stanene nanoribbon under the perpendicular electric field is theoretically investigated.  Optical response of spin optoelectronic nanodevice in the presence of  linearly polarized illumination is calculated by means of the self-consistent nonequilibrium Green’s function approach  together with the tight-binding model. In antiferromagnetic $ SZNR $, twofold spin degeneracy of spin-resolved energy levels is split due to combination effect of the vertical electric field and the large spin-orbit interaction in stanene lattice. Also, robustness of the spin-polarized photocurrent has been demonstrated when the electric field is applied on the two edges.  Interestingly, the spin-polarized current is generated in the wide wavelength region of incident light in the presence of the edge potential. Besides, the results indicate that band bending enhances spin-polarization in device. In particular, it is shown that in the presence of the narrow edge potential, by varying antiferromagnetic exchange field, spin-semiconducting behavior is obtained in stanene nanoribbon.  This behavior arises as a consequence of the large spin-orbit coupling in stanene lattice. Furthermore, it is found that  the spin-resolved photocurrent can be engineered by external strain and  nearly full optical spin-filtering is obtained in antiferromagnetic stanene lattice under the combining effect of strain and narrow edge potential. The obtained results in this study may be useful to develop stanene-based nanodevices such as spin-photo detectors, spin photodiodes and generation of full spin-filtering.  
      
 \section*{MODEL AND METHOD}
 
 The proposed spin optoelectronic device is designed based on $ ZSNR $ in the presence of some external field, such as normal electric field, antiferromagnetic field and linearly polarized light field. The performed simulations are divided into two self-consistent computations, in which the first part is based on calculating spin-dependent electron features of $ ZSNR $ and the second part taking into account the quantum transport equation of the interaction of light with  matter by employing the $ NEGF $ approach. It is worthy to mention that due to the absence of impurity or the electron-phonon interaction in the present study, spin-flip mechanisms are neglected~\cite{kurpas2019spin}. Consequently,  transport equations have been solved for spin-up and spin-down components, individually.
 
 The total Hamiltonian of the proposed nanodevice is divided as follows:
 
 \begin{equation}
 H_{T}=H_{L}+H_{R}+H_{C}+H_{CL}+H_{CR}.
 \end{equation}
 
 Where the first two contributions in Eq.(1) are the Hamiltonian of semi-infinite left and right leads, respectively. $ H_{LC} $ and $ H_{RC} $ describe coupling
 between the scattering region and the left and right leads. $ H_{C} $ represents the Hamiltonian of scattering region:
 \begin{equation}
 H_{C}=H_{0}+H_{e\gamma}.
 \end{equation}
 
 $ H_{0} $ is the tight-binding model in the presence of antiferromagnetic field and the normal electric field as follows~\cite{ezawa2012valley, shakouri2015tunable}:
 
 \begin{eqnarray}
 H_{0}&=&-t\sum_{\langle\,i,j \rangle , \alpha}c^{\dagger}_{i\alpha}c_{j\alpha}+i\frac{\lambda_{so}}{3\sqrt{3}}\sum_{\langle\langle\,i,j \rangle , \alpha,\beta}\nu_{ij}c^{\dagger}_{i\alpha}\sigma^{z}_{\alpha\beta}c_{j\beta}
 \nonumber\\
 &&+e\ell E_{z}\sum_{i,\alpha}\xi_{i}c^{\dagger}_{i\alpha}c_{i\alpha}+M_{AF}\sum_{i,\alpha,\beta}\xi_{i}c^{\dagger}_{i\alpha}\sigma^{z}_{\alpha\beta}c_{j\beta}.
 \end{eqnarray}
 
 Where $ t $ is the usual nearest-neighbor hopping in the scattering region and its value is equal to $ 1.3\: eV $. $ c_{i\alpha}(c^{\dagger}_{i\alpha}) $ annihilates (creates) an electron with spin polarization $ \alpha $ at atom $ i $, and $ <i,j> $\:($ <<i,j >> $)  represents the sum over nearest (next-nearest) Sn-Sn pairs. The second term in Eq.(3) accounts for the effective spin orbit coupling parameter with a coupling strength of $ \lambda_{so}=100\:meV $. $ \vec{\sigma}=(\sigma_{x},\sigma_{y}, \sigma_{z}) $ is the Pauli’s spin matrix. $ \nu_{ij}=-1(+1) $  for clockwise (anticlockwise) next-nearest-neighboring hopping.  The third contribution denotes the effect of  a perpendicular electric field $ E_{z} $. Also, $\ell $ is the buckling height and  $ \xi_{i} $ is equal to $ +1(-1) $ for the upper (lower) sublattice. The last term is related to antiferromagnetic  exchange field~\cite{lu2019spin, luo2019antiferromagnetic}. $ H_{e\gamma}=\frac{e}{m_{e}}\,\vec{A}\,.\,\vec{P} $ indicates the electron-photon interaction and it is considered
 as the first-order perturbation Hamiltonian. $ m_{e} $ is the mass of electron, $ \vec{A} $ and $ \vec{P} $ represent the time-dependent electromagnetic vector potential and the momentum of the electron, respectively~\cite{henrickson2002nonequilibrium, aeberhard2008microscopic}.
 
 After computing the Hamiltonian of the scattering region and the left and right leads, the retarded Green’s function of the nanodevice, in the presence of light radiation, can be written as:
 \begin{eqnarray}\label{trepresent}
 &&G_{\sigma}(E)=[(E+i\eta)I-H_{C,\sigma}-\Sigma_{T,\sigma}]^{-1},
 \end{eqnarray}
 where
 \begin{eqnarray}\label{trepresent}
 &&\Sigma_{T,\sigma}=\Sigma_{L,\sigma}+\Sigma_{R,\sigma}+\Sigma_{\gamma,\sigma}.
 \end{eqnarray}
 In Eq.(4), $ \eta $  and $ I $ are infinitesimal broadening and identity matrix, respectively. $ \Sigma_{L(R),\sigma} $ is the retarded self-energy due to the presence of the left and right leads. While the self-energies of the left and right contacts were computed by the Sancho iterative approach~\cite{sancho1985highly, li2008quantum}, the electron-photon scattering is included as self-energy term. In Eq.(5), $ \Sigma_{\gamma,\sigma} $ is the self-energy of the electron-photon interaction  which is expressed as:
 
 \begin{eqnarray}\label{Fockspace}
 \Sigma_{\gamma,\sigma}=\frac{-i}{2}[\Sigma_{\gamma,\sigma}^{\textless}(E)-\Sigma_{\gamma,\sigma}^{\textgreater}(E)].
 \end{eqnarray}

 $\Sigma_{\gamma,\sigma}^{\textless} $ and $ \Sigma_{\gamma,\sigma}^{\textgreater} $ are the lesser and greater self-energies of the electron-photon scattering which are given as follows:
 
 \begin{eqnarray}\label{Fockspace}
 \Sigma_{\gamma,\sigma}^{\gtrless}(E)&=&(N_{\gamma}+1)M^{\gamma}G_{\sigma}^{\gtrless}(E^{\mp})M^{\gamma}
 \nonumber\\
 &&+N_{\gamma}M^{\gamma}G_{\sigma}^{\gtrless}(E^{\pm})M^{\gamma}.
 \end{eqnarray}
 
 In the above equation, $ E^{\pm}=E\pm\hslash\,\omega $ and $ N_{\gamma} $ exhibits the number of photon with energy $ \hslash\,\omega $~\cite{aeberhard2011quantum}. $ M^{\gamma} $ is the electron-photon interaction arising from the perturbation Hamiltonian $ H_{e\gamma} $. Each element of $ M^{\gamma} $ is given by:
 \begin{eqnarray}
 M_{lm}^{\gamma}=\langle\,l|\vec{P}.\,A_{0}\,\hat{e_{p}}|\,m\rangle. 
 \end{eqnarray}
 $ A_{0} $ is the amplitude of electromagnetic vector potential and it’s direction is determined by the light polarization ($ \hat{e_{p}} $). $ G_{\sigma}^{\textless} $ is the electron correlation function~\cite{keldysh1965diagram}:
 \begin{eqnarray}
 G_{\sigma}^{\textless}(E)&=&G_{\sigma}(E)[\Gamma_{L,\sigma}(E)\,f_{L}(E)+\Gamma_{R,\sigma}(E)\,f_{R}(E)
 \nonumber\\
 &&+\Sigma_{\gamma,\sigma}^{\textless}]G_{\sigma}^{\dagger}(E)
 \end{eqnarray}
 and the hole correlation function is:
 \begin{eqnarray}
 G_{\sigma}^{\textgreater}(E)&=&G_{\sigma}(E)[\Gamma_{L,\sigma}(E)\,f_{L}(E)+\Gamma_{R,\sigma}(E)\,f_{R}(E)
 \nonumber\\
 &&+\Sigma_{\gamma,\sigma}^{\textgreater}]G_{\sigma}^{\dagger}(E).
 \end{eqnarray}
 In Eqs.(9) and (10), $ f_{L}(_{R}) $ is the left (right) Fermi-Dirac function. $ \Gamma_{L(R),\sigma}= i(\Sigma_{L(R),\sigma}-\Sigma_{L(R),\sigma}^{\dagger})$, represents the broadening functions of the left (right) electrode. $\Sigma_{\gamma,\sigma} $ is determined  self-consistently by the iteration method.  Once convergence is obtained, one can calculate the spin photocurrent across the system by~\cite{anantram2008modeling}:
 \begin{eqnarray}
 I_{\gamma,\sigma}&=&\frac{2e}{\hslash}\int\,\frac{dE}{2\pi}\,Tr[G_{(1,1),\sigma}^{\textgreater}(E)\Gamma_{L,\sigma}(E)\,f_{L}(E)
 \nonumber\\
 &&-G_{(1,1),\sigma}^{\textless}(E)\Gamma_{L,\sigma}(E)\,(1-f_{L}(E))].
 \end{eqnarray}
 Where $ G_{(1,1),\sigma}^{\textgreater}(G_{(1,1),\sigma}^{\textless})$ is the first block of the hole (electron) correlation function~\cite{tao2018h}.
 
 Spin-dependent quantum efficiency can be written as:
 
 \begin{eqnarray}\label{Fockspace}
 \eta_{\sigma}=\frac{E_{ph}\,I_{ph,\sigma}}{e\,A_{D}\,I_{w}}\times100\: \%\:(\sigma=up,\:down).
 \end{eqnarray}
 
 Where $ A_{D}=L_{ch}\,W_{ch} $ and $E_{ph}$ are cross section of central channel and photon energy, respectively. Also, spin polarization is defined as $ SP(\%)=(|I_{ph,\sigma}|-|I_{ph,-\sigma}|)/(|I_{ph,\sigma}|+|I_{ph,-\sigma}|)\times100\: \% $.\\

  %To clarify the edge potential on the electronic structure of ZSNR, the band structure when the electric field is applied to  the whole area of the nanoribbon is plotted  by black curves.

%\nocite{*}
\section*{DATA AVAILABILITY}
The data and codes that support the findings of this study are available from the authors upon reasonable request.


\section*{acknowledgements}
This research is supported by Post-doctoral research grant of Azarbaijan Shahid Madani University (218/D/28672).
\section*{author contributions}
F.R. performed computations of the work and written the manuscript, A.P. reviewed and analyzed the results.
\section*{competing interests}
The authors declare no competing interests.

\newpage
%\begin{figure*}
%	\includegraphics{1(a).eps}% Here is how to import EPS art
%	\includegraphics{1(c).eps}% Here is how to import EPS art
%	\caption{\label{fig:wide}Use the figure* environment to get a wide
%		figure that spans the page in \texttt{twocolumn} formatting.}
%\end{figure*}

\newpage

\newpage

\newpage

\newpage

\newpage

\newpage

\newpage

\newpage
\newpage

\newpage

\end{document}